\documentclass[aps,prc,twocolumn,showpacs,preprintnumbers,nofootinbib,float,superscriptaddress]{revtex4-1}

\usepackage{epsfig}
 
\usepackage[dvipsnames]{xcolor}
\usepackage{float,amsmath,amssymb}
\usepackage{graphicx}
\usepackage{url}
\usepackage{bbold}
\usepackage[utf8]{inputenc}
\usepackage{physics}
\usepackage{float}
\newcommand{\br}{\mathbf{r}}

\newcommand{\xpom}{x_{I\hspace{-0.3em}P}}
\newcommand{\gev}{\mathrm{GeV}}

\newcommand{\xbj}{{x}}
\newcommand{\ygap}{Y_{\rm gap}}

\newcommand{\qs}{{Q_\mathrm{s}}}

\newcommand{\lqcd}{\Lambda_{\mathrm{QCD}}}
\newcommand{\as}{{\alpha_{\mathrm{s}}}}
\newcommand{\rt}{{\mathbf{r}}}
\newcommand{\ra}{{\mathbf r}_1}
\newcommand{\rb}{{\mathbf r}_2}
\newcommand{\bt}{{\mathbf{b}}}
\newcommand{\ba}{{\mathbf{b}}_1}
\newcommand{\bb}{{\mathbf{b}}_2}

\newcommand{\Deltat}{{\boldsymbol{\Delta}_\perp}}

\newcommand{\nc}{{N_\mathrm{c}}}

\usepackage[breaklinks,colorlinks,citecolor=citcolor,urlcolor=blue,linkcolor=lcolor]{hyperref}
\definecolor{lcolor}{rgb}{0.5,0,0}
\definecolor{citcolor}{rgb}{0,0.3,0.0}
\usepackage[capitalise]{cleveref}
\usepackage{todonotes}

\begin{document}

\title{Rapidity gap distribution of diffractive small-$\xpom$ events at HERA and at the EIC}

\author{Tuomas Lappi}
\author{Anh Dung Le}
\author{Heikki Mäntysaari}

\affiliation{Department of Physics, University of Jyväskylä, P.O. Box 35, 40014 University of Jyväaskylä, Finland}
\affiliation{Helsinki Institute of Physics, P.O. Box 64, 00014 University of Helsinki, Finland}

\begin{abstract}
   We use the Kovchegov-Levin equation to resum contributions of large invariant mass diffractive final states to diffractive structure functions in the dipole picture of deep inelastic scattering. 
For protons we use a (modified) McLerran-Venugopalan model  as the initial condition for the evolution, with free parameters obtained from fits to the HERA inclusive data.  We obtain an adequate agreement to the HERA diffractive data in the moderately high-mass regimes
   when the proton density profile is fitted to the diffractive structure function data in the low-mass region.
The HERA data is found to prefer a proton shape that is steeper than a Gaussian.
The initial conditions are generalized to the nuclear case using the optical Glauber model. 
   Strong nuclear modification effects are predicted in diffractive scattering off a nuclear target in  kinematics  accessible at the future Electron-Ion collider. In particular, the Kovchegov-Levin evolution has a strong effect on the $Q^2$-dependence of the diffractive cross section.
\end{abstract}

\maketitle

\section{Introduction}

Diffractive processes in deeply inelastic electron-hadron scattering (DIS) with no net color charge transfer are powerful in probing the high-energy structure of protons and nuclei. The color singlet exchange requires, at lowest order in perturbative QCD, two gluons to be exchanged,  rendering  diffractive cross sections more sensitive to the gluonic content of the target  than inclusive ones. Consequently, high-energy diffraction can provide clear indications for gluon saturation effects, which are expected to occur in the regime of small longitudinal momentum fraction $x$ due to non-linear QCD dynamics. 

The Color Glass Condensate (CGC) effective theory provides a convenient framework to describe scattering processes at high energy~\cite{Gelis:2010nm}. Instead of (inclusive or diffractive) parton distribution functions, the target structure is described in terms of Wilson lines that describe an eikonal propagation of projectile partons in the target color field. In DIS, this CGC approach is frequently complemented with the dipole picture~\cite{Kopeliovich:1981pz,Mueller:1989st, Nikolaev:1990ja} in a frame where the virtual photon mediating the interaction has a large longitudinal momentum, so that its $|q\Bar{q}\rangle$ Fock state (and possibly $|q\Bar{q} g\rangle, |q\Bar{q} g g\rangle \dots$) has a long lifetime compared to the typical timescale of the interaction. The dipole model is particularly suitable to the study of gluon saturation. A particular advantage of this CGC + dipole picture is that it provides a common theoretical framework to incorporate the description of both inclusive and diffractive scattering processes in terms of the same degrees of freedom.

The CGC + dipole formalism has been widely employed in studying the diffractive dissociation of the photon off both protons and nuclei~\cite{Munier:2003zb, Marquet:2007nf,Kowalski:2008sa, Kugeratski:2005ck, Cazaroto:2008iy, Bendova:2020hkp, Beuf:2022kyp, Kovchegov:1999ji,Kovchegov:2011aa, Kovner:2006ge, Lublinsky:2014bma, Levin:2001pr,Levin:2001yv,Levin:2002fj,Contreras:2018adl,Le:2021afn,Hatta:2006hs, Hatta:2006zz}.
One of the main advantages of this framework is that  saturation effects appear
naturally, consistently in both diffractive and inclusive cross sections.  
In practice, starting from the work in~\cite{Golec-Biernat:1999qor,Golec-Biernat:2001gyl}, two quantum Fock state components $|q\Bar{q}\rangle$ at leading order and $|q\Bar{q}g\rangle$ (part of the next-to-leading order contribution) in approximative kinematics  have been  considered in order to compare to the available HERA diffractive data~\cite{Munier:2003zb,Marquet:2007nf,Kowalski:2008sa} as well as to make some predictions for future experiments~\cite{Kugeratski:2005ck, Cazaroto:2008iy, Kowalski:2008sa, Bendova:2020hkp}. 

Recently, there has been a rapid progress towards next-to-leading  (NLO) order accuracy.
Developments that are necessary to achieve the NLO level in theoretical calculations include the tree-level diffractive $q\Bar{q} g$ production in exact kinematics~\cite{Beuf:2022kyp} and loop corrections to the virtual photon wave functions describing the $\gamma^*\to q\Bar{q}$ splitting~\cite{Hanninen:2017ddy,Beuf:2021srj}.
In another aspect, there have been also attempts to resum soft gluon contributions in the regime of high-mass diffraction~\cite{Kovchegov:1999ji,Kovchegov:2011aa, Kovner:2006ge, Lublinsky:2014bma,Levin:2001pr,Levin:2001yv,Levin:2002fj,Contreras:2018adl,Le:2021afn}. Such improvements in precision are particularly important for phenomenological studies related to future DIS facilities such as the future Electron-Ion Collider (EIC)~\cite{AbdulKhalek:2021gbh} and the LHeC/FCC-he~\cite{LHeC:2020van}. 
 These future facilities are expected to provide very precise data for diffractive observables over a wide kinematical domain. In particular the first measurements for nuclear diffractive structure functions will be performed at the EIC in the 2030s.
These measurements  with nuclear targets are especially of interest as they are highly sensitive to the gluon saturation effects~\cite{Aschenauer:2017jsk,Accardi:2012qut}, which are strongly enhanced by either going to smaller $x$ or heavier nuclei.  

In this work we focus on diffractive DIS in the region where the mass of the diffractively produced system is large, which requires the resummation of soft-gluon contributions by the means of the Kovchegov-Levin equation~\cite{Kovchegov:1999ji,Kovchegov:2011aa,Kovchegov:2012mbw, Lublinsky:2014bma}. This perturbative evolution equation requires non-perturbative input sensitive to the proton structure at moderately small $x$, which can be constrained by HERA inclusive structure function data (see also Ref.~\cite{Dumitru:2023sjd} for a complementary approach starting from the proton large-$x$ structure). The predictions for high-mass diffraction in electron-proton DIS at HERA and electron-nucleus DIS at the EIC are genuine predictions, once the initial condition for the Balitsky-Kovchegov (BK) evolution~\cite{Balitsky:1995ub,Kovchegov:1999yj} of the 
  dipole amplitude has been fit to inclusive cross section data.  The only additional free parameter in the calculation is the spatial density profile of the proton,  whose functional form is not probed in inclusive structure function measurements. Here we constrain this impact parameter profile with the HERA diffractive structure function data in the low-mass regime.

The paper is organized as follows. In the next section, we review the dipole picture of (diffractive) deep-inelastic scattering and the evolution equations in the CGC approach for both inclusive and diffractive processes. Both low-mass and high-mass approaches for diffraction are discussed for a more complete treatment. The application of our setup to the HERA diffractive data is then presented in~\cref{sec:hera_comparisons}. In \cref{sec:EIC-prediction} we  make  predictions for nuclear diffraction in kinematics accessibe at the future EIC. We finally draw some concluding remarks in~\cref{sec:conclusions}.

\section{Diffractive deep inelastic scattering in the dipole picture}

\subsection{Dipole picture and diffractive observables}

\begin{figure} [tb]
    \centering
    \includegraphics[width=\columnwidth,height=\textheight,keepaspectratio]{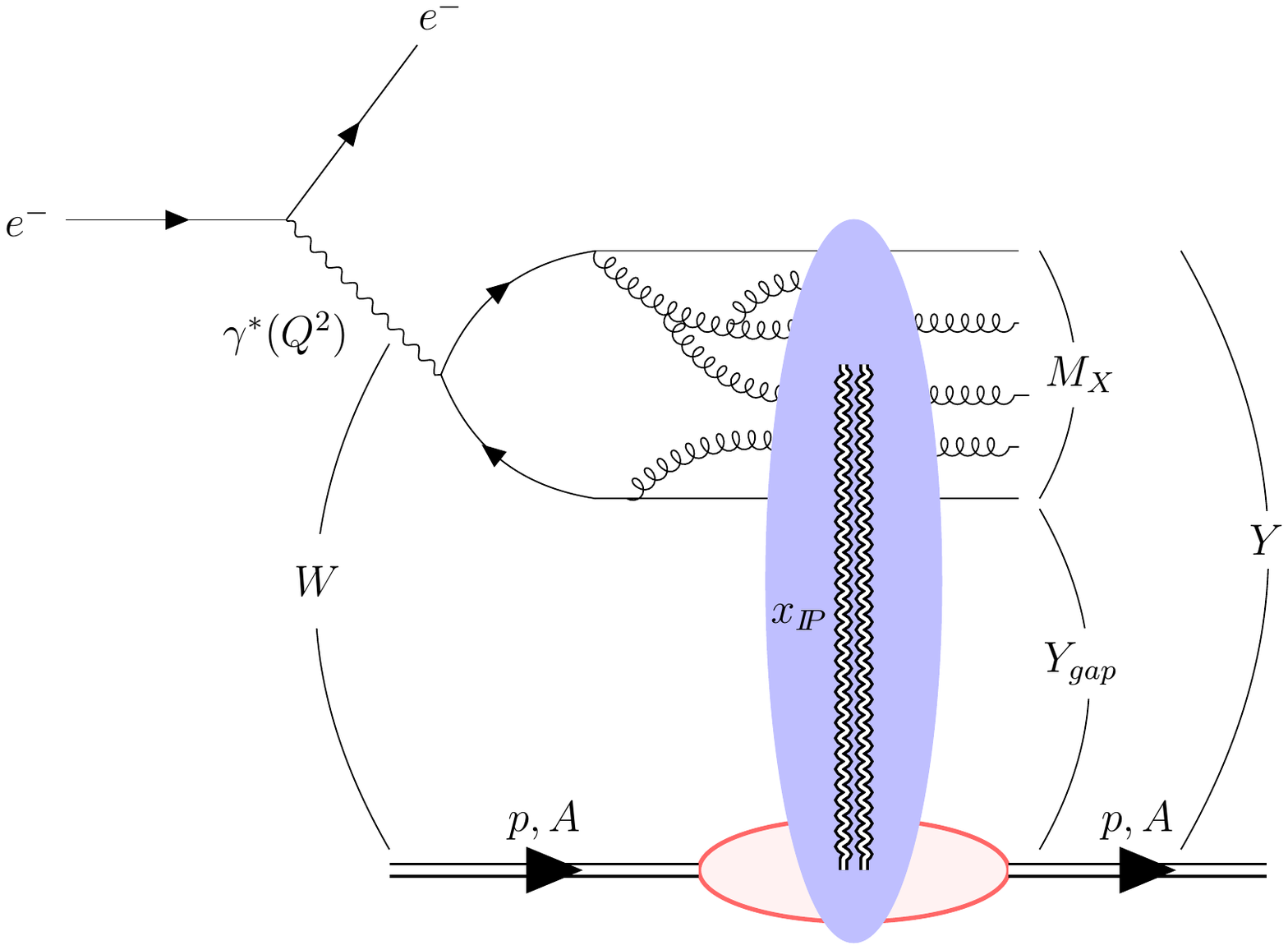}
    \caption{Diffractive dissociation in the dipole picture. The rapidity gap $\ygap$ in the final state is due to the pomeron exchange (represented by the double wavy line) taking a momentum fraction $\xpom$ of the hadron. Relevant kinematic variables described in the text are also shown.}
    \label{fig:DDIS_illustration}
\end{figure}

Within the single-photon approximation, the deep inelastic interaction between the electron and a hadron is mediated by a   photon of virtuality $Q^2$. At high center-of-mass energy $W$ of the photon-hadron sub-process, it is convenient to go to a reference frame where the photon has a large longitudinal momentum. In this frame, its coherence length in the longitudinal direction is larger than the size of the hadronic target. Hence, if the photon branches into a quark-antiquark dipole, this quantum fluctuation will occur long before traversing the target and, to a good approximation, the transverse size of the resulting dipole will remain unchanged during the interaction (see~\cref{fig:DDIS_illustration}). Consequently, the dipole-proton scattering amplitude becomes a good degree of freedom to describe (both inclusive and diffractive) scattering processes at high energy.

In the diffractive dissociation process of interest, the diffractively produced system of invariant mass $M_X$ in the final state results  from the fragmentation of the dipole (possibly dressed by other partons from higher-order quantum fluctuations), while the target hadron remains intact. We only consider coherent diffraction in this work. An experimental signature of such a diffractive scattering is a rapidity gap $\ygap \le Y$, with Y being the total relative rapidity, between the diffractively produced system and the outgoing hadron, as illustrated in~\cref{fig:DDIS_illustration}. In the theoretical point of view, this rapidity gap is due to the exchange of a color-singlet C-even pomeron in the $t$ channel.
When the momentum transfer is integrated out, the diffractive scattering process can be completely characterized by three invariants $Q^2$, $W$ and $M_X^2$. Alternatively, one can use instead the variables $\xpom$, $\beta$ and $Q^2$, where 
\begin{equation}
\label{eq:defbeta}
\beta = \frac{Q^2}{Q^2+M_X^2}
\end{equation}
 and 
\begin{equation}
\label{eq:defxpom}
\xpom = \frac{Q^2+M_X^2}{Q^2+W^2}.
\end{equation}
 In the pomeron exchange picture, $\xpom$ can be interpreted as the fraction of the target longitudinal momentum carried by the pomeron (in the infinite momentum frame) and $\beta$ is the momentum fraction of the pomeron carried by the struck parton. Note that these are related to the Bjorken variable as $x=\xpom\beta$. By definition, the rapidity variables are linked to these momentum fractions as $Y = \ln (1/x)$ and $\ygap=\ln (1/\xpom)$.

Before going to more details of the formulation, let us define the diffractive observables of interest for the current analysis. The experimentally determined diffractive structure functions $F_{2,L}^{D(3)}$ are related to the diffractive virtual photon-hadron cross sections as
\begin{equation}
    \xpom F_{L}^{D(3)} = \frac{Q^2}{4\pi^2\alpha_{em}} \frac{\dd \sigma_{D\ (L)}^{\gamma^*h}}{\dd \ln (1/\beta)},
\end{equation}
and
\begin{equation}
\label{eq:f2D_def}
    \xpom F_{2}^{D(3)} = \frac{Q^2}{4\pi^2\alpha_{em}} \left( \frac{\dd\sigma_{D\ (T)}^{\gamma^*h}}{\dd\ln (1/\beta)} + \frac{\dd\sigma_{D\ (L)}^{\gamma^*h}}{\dd \ln (1/\beta)}\right),
\end{equation}
where  $T$ and $L$ refer to the polarization state of the virtual photon.
%
The most precise diffractive cross section measurements from HERA~\cite{H1:2012xlc} are reported as a reduced diffractive cross section defined as 
\begin{equation}    \label{eq:red_cross_section}
    \sigma^{D(3)}_\mathrm{red} = F_2^{D(3)} - \frac{y^2}{1 + (1-y)^2} F_L^{D(3)},
\end{equation}
where $y = Q^2/(xs)$ is the inelasticity, and $\sqrt{s}$ is the center-of-mass energy of the electron-proton scattering. The superscript ``(3)'' in the above formulae indicates that the relevant observables depend on three invariants, as mentioned above: in this work we only consider the case where the cross section is integrated over the squared momentum transfer $t$. 
The diffractive cross section can also be expressed in terms of the mass of the diffractive system as
\begin{equation}
    \label{eq:mass_spectrum_def}
    \frac{\dd\sigma_D^{\gamma^*h}}{\dd{M_X}} = \frac{2M_X}{Q^2 + M_X^2}\left( \frac{\dd\sigma_{D\ (T)}^{\gamma^*h}}{\dd\ln (1/\beta)} + \frac{\dd\sigma_{D\ (L)}^{\gamma^*h}}{\dd\ln (1/\beta)}\right).
\end{equation}

The current investigation employs two different approaches to calculate diffractive cross sections. In the large $\beta$  (small $M_X^2$) regime we use explicit results computed considering the $q\Bar{q}$ and $q\Bar{q} g$ components of the virtual photon ($q\Bar{q} g$ only in the high-$Q^2$ limit), which have been extensively used in the literature, see e.g. Ref.~\cite{Kowalski:2008sa}. We use these results as a baseline to fix the one remaining free parameter related to the proton spatial density profile as discussed in detail below. Then with no free parameters we calculate diffractive structure functions at small $\beta$ (large $M_X^2$) by solving the Kovchegov-Levin evolution equation which resums contributions from dipole states dressed by soft gluons. These two approaches are reviewed below.

\subsection{High-energy evolution and inclusive scattering}

In the framework of the dipole picture, the strong interaction dynamics is encoded in the forward dipole-target elastic scattering amplitude $N(\rt, Y; \bt)$, where $\rt$ is the transverse size of the dipole and $\bt$ is the dipole-target impact parameter.
At a large number of colors $\nc$, the energy (or rapidity $Y$) dependence of the dipole amplitude is given by the Balitsky-Kovchegov (BK) equation~\cite{Balitsky:1995ub,Kovchegov:1999yj}
\begin{multline}    \label{eq:BK_eq}
    \partial_Y N(\rt,Y;\bt) = \int \dd[2]{\ra} \mathcal{K}({\rt}, \ra, \rb) \left[ N(\ra,Y;\ba)  \right. \\
    \left.+ N(\rb,Y;\bb) - N(\rt,Y;\bt) - N(\ra,Y;\ba)N(\rb,Y;\bb)\right], 
\end{multline}
where $\rb = \rt - \ra$, $\ba = \bt - (\rb/2)$ and $\bb = \bt + (\ra/2)$. The kernel $\mathcal{K}({\rt}, \ra, \rb)$ is related to the probability amplitude, at large $\nc$, for emitting a soft gluon at a point in the transverse plane characterized by two vectors $\ra$ and $\rb$ satisfying the triangular relation $\rt = \ra + \rb$ from the initial dipole.  In this work we use the leading order BK equation~\eqref{eq:BK_eq}, but  include the running coupling corrections to the evolution by adopting the Balitsky running-coupling prescription~\cite{Balitsky:2006wa}, which reads
\begin{multline}    \label{eq:rc_kernel}
    \mathcal{K}^{rc} ({\rt}, \ra, \rb) = \frac{N_c\alpha_s(r^2)}{2\pi^2} \left[ \frac{r^2}{r_1^2r_2^2} + 
    \frac{1}{r_1^2}\left(\frac{\alpha_s(r_1^2)}{\alpha_s(r_2^2)}-1\right) \right.\\ 
    \left. + \frac{1}{r_2^2}\left(\frac{\alpha_s(r_2^2)}{\alpha_s(r_1^2)}-1\right)\right].
\end{multline}
The strong coupling constant in coordinate space is taken as
\begin{equation}    \label{eq:running_coupling}
    \alpha_s(r^2) = \frac{12\pi}{(33-2N_f)\ln \frac{4C^2}{r^2\lqcd^2}}. 
\end{equation}
To avoid the Landau pole, the running coupling is frozen at the value $\alpha_s^\mathrm{fr} = 0.7$ for $r^2>r_\mathrm{fr}^2$, where $r_\mathrm{fr}^2$ solves $\alpha_s(r_\mathrm{fr}^2) = \alpha_s^\mathrm{fr}$. The constant $C^2$ in the above formula accounts for the uncertainty when transforming from momentum space to coordinate space. From theoretical considerations~\cite{Balitsky:2006wa,Kovchegov:2006vj} it should have the value $e^{-2\gamma_E}$. 
In practice, however, the running coupling scale in coordinate space is taken as a free parameter that can absorb some dominant higher order effects that would slow down the evolution. The non-perturbative initial condition for the BK equation and the value of $C^2$ are obtained from a fit to proton inclusive structure function data e.g. in Refs.~\cite{Lappi:2013zma,Albacete:2010sy} (see also recent fits at next-to-leading order accuracy~\cite{Beuf:2020dxl,Hanninen:2022gje} that however can not be used in the leading order calculation presented here).
In this work we use the fits reported in Ref.~\cite{Lappi:2013zma}, and consequently adopt the same setup and work with only light quarks ($N_f=3, m_f=140\,\mathrm{MeV}$). 
The considered fits initialize the BK evolution at rapidity
$Y_\mathrm{min} \equiv \ln (1/x_\mathrm{init})$ with $x_\mathrm{init} = 0.01$. The initial condition for the BK equation at this starting point is discussed in more detail in~\cref{sec:hera_comparisons,sec:EIC-prediction} for the scattering off protons and nuclei, respectively. 

Given the forward dipole elastic amplitude $N$, the total (inclusive) dipole-target cross section $\sigma^{q\Bar{q}h}_\mathrm{tot}$ can be computed straightforwardly using the optical theorem. Integrating out the $\bt$-dependence one obtains
\begin{equation}
    \label{eq:dipole_tot_cs}
    \sigma^{q\Bar{q}h}_\mathrm{tot} (\rt, Y) = \int \dd[2]{\bt} \ 2N(\rt,Y,\bt). 
\end{equation}
Convoluting with the photon impact factor, we eventually obtain the total (inclusive) photon-target  cross section
\begin{equation}
\label{eq:photon_tot_cs}
\begin{aligned}
    \sigma^{\gamma^{*}h}_\mathrm{tot} (Q^2, Y) = \sum_f \int \dd[2]{\rt} \int_0^1 \dd{z} & \big\lvert \psi^{\gamma^*\to f\Bar{f}} (\rt,z,Q^2)\big\rvert^2_{T+L} \\
    & \qquad\quad\times \sigma^{q\Bar{q}h}_\mathrm{tot} (\rt, Y),
\end{aligned}
\end{equation}
where the photon wave functions $\psi_{T,L}^{\gamma^*\to f\Bar{f}}$ can be computed from QED using light cone perturbation theory~\cite{Kovchegov:2012mbw}. 
Only light quark flavors are included in this work.

\subsection{Diffraction at large and medium $\beta$}

Now we turn to the calculation of diffractive observables. For medium to large values of $\beta$, it is enough to consider only the two lowest order (in $\as$) partonic states of the virtual photon, $|q\Bar{q}\rangle$ and $|q\Bar{q}g\rangle$. We quote here the well-known results for these contributions studied e.g. in Ref.~\cite{Kowalski:2008sa}. The $q\Bar q$ contribution dominates at large $\beta \gtrsim 0.5$, and the diffractive structure functions for transversely and longitudinally polarized virtual photons read
\begin{multline}
\label{eq:gbw_qq_T}
    \xpom F^{D(3)}_{q\Bar{q},T} = \frac{N_c Q^4}{16\pi^3 \beta}\sum_f e_f^2\int\limits_{z_0}^{1/2} \dd{z} z (1-z) \\ \left[\epsilon^2\left(z^2+(1-z)^2\right)\Phi_1 + m_f^2\Phi_0\right], 
\end{multline}
and
\begin{multline}
\label{eq:gbw_qq_L}
    \xpom F^{D(3)}_{q\Bar{q},L} = \frac{N_c Q^6}{4\pi^3 \beta}\sum_f e_f^2\int\limits_{z_0}^{1/2} \dd{z} z^3 (1-z)^3 \Phi_0.
\end{multline}
Here we have used the following auxiliary function 
\begin{equation}
    \label{eq:auxiliary_functions}
    \Phi_n = \int \dd[2]{\bt} \left[ 2\int\limits_0^{\infty} \dd{r} r K_n(\epsilon r) J_n(\kappa r) N(\rt,\ygap;\bt) \right]^2,
\end{equation}
with $\epsilon^2 = z(1-z)Q^2 + m_f^2$, $\kappa^2 = z(1-z)M_X^2 - m_f^2$ and $z_0 = \left(1-\sqrt{1-4m_f^2/M_X^2}\right)/2$.

Toward smaller $\beta\lesssim 0.5$ the contribution from one gluon emission becomes important. The diffractive $q\Bar q g$ production is known in exact kinematics~\cite{Beuf:2022kyp}, but in phenomenological applications so far only the so called Wusthoff result~\cite{Wusthoff:1997fz} obtained in the large $Q^2$ limit  
has been used. In that limit the transverse polarization dominates, by the means of the $\ln Q^2$ enhancement compared to the longitudinal one. Furthermore the $q\Bar{q}g$ system can be treated as an effective gluon dipole. The resulting contribution to the diffractive structure function reads
\begin{multline}
    \label{eq:gbw_qqg}
    \xpom F^{D(3)}_{q\Bar{q}g,T} = \frac{\alpha_s(Q^2)\beta}{8\pi^4} \sum_f e_f^2 \int \dd[2]{\bt} \int\limits_{0}^{Q^2} \dd{k^2} \int\limits_{\beta}^{1} \dd z \left\{ \vphantom{\left[2\int\limits_0^{\infty} \right]^2} \right.\\
    \left. k^4\ln\frac{Q^2}{k^2}\left[\left(1-\frac{\beta}{z}\right)^2 + \left(\frac{\beta}{z}\right)^2 \right] \right.\\
    \left. \times \left[2\int\limits_0^{\infty} \dd{r} r K_2(\sqrt{z}k r) J_2(\sqrt{1-z} kr)\Tilde{N}(\rt,Y_{gap};\bt)\right]^2\right\},  
\end{multline}
with $\Tilde{N} = 2N-N^2$ representing the dipole-target amplitude in the adjoint representation. Here we choose to evaluate the strong coupling constant $\as$ at the scale $Q^2$.

Note that in~\cref{eq:gbw_qq_T,eq:gbw_qq_L,eq:auxiliary_functions,eq:gbw_qqg} the dipole-target amplitudes are evaluated at the rapidity $\ygap$, since this low-mass diffaction can be treated as a quasi-elastic scattering process with $Y \approx \ygap$ and $F^{D(3)}_{q\Bar{q}} \sim N^2\ ({\rm\ or\ }F^{D(3)}_{q\Bar{q}g} \sim\Tilde{N}^2)$. Recall that since we start the BK evolution at $Y_\mathrm{min} \equiv \ln (1/x_\mathrm{init})$, then $\ygap \ge Y_\mathrm{min}$ or $\xpom \le x_\mathrm{init}$.  We will refer to these low-mass contributions as the GBW result\footnote{In their pioneering works~\cite{Golec-Biernat:1999qor,Golec-Biernat:2001gyl,Wusthoff:1997fz}, Golec-Biernat and W{\"u}sthoff (GBW) used their saturation model for the dipole-target interaction instead of the BK-evolved dipole amplitudes used in the current study.} hereafter. 

\subsection{Diffraction at small-$\beta$ and the Kovchegov-Levin evolution equation}
\label{sec:kl}

At small $\beta$, higher-order gluonic states are essential, and it is necessary to resum soft gluon emissions to all orders. At large $N_c$, this resummation can be done by using the Kovchegov-Levin (KL) evolution equation.
Denoting the diffractive dipole-target cross section at fixed impact parameter $\bt$ and with a \emph{minimal} rapidity gap 
$Y_0$  by  $N_D(\rt,Y,Y_{0};\bt)$, the KL equation reads\footnote{The KL equation is known at NLO, see Ref.~\cite{Lublinsky:2014bma}, which has the same form as the NLO BK equation~\cite{Balitsky:2008zza}. Here we restrict ourselves to only the running-coupling correction consistently with our leading-log setup.}~\cite{Kovchegov:1999ji,Kovchegov:2011aa,Kovchegov:2012mbw} 
\begin{multline}    \label{eq:KL_eq}
     \partial_Y N_D(\rt,Y,Y_0;\bt) =  \int \dd[2]{{\bf r}_1} \mathcal{K}({\bf r}, {\bf r}_1, {\bf r}_2) 
     \left[ N_D(\ra,Y,Y_0;\ba) \right. \\
     \left. + N_D(\rb,Y,Y_0;\bb)- N_D(\rt,Y,Y_0;\bt) \right. \\
    \left. + N_D(\ra,Y,Y_0;\ba)N_D(\rb,Y,Y_0;\bb) \right. \\
    \left. + 2N(\ra,Y;\ba)N(\rb,Y;\bb) \right.  \\
    \left. - 2N_D(\ra,Y,Y_0;\ba)N(\rb,Y;\bb) \right. \\
    \left. - 2N(\ra,Y;\ba)N_D(\rb,Y,Y_0;\bb) \right].
\end{multline}
The initial condition for the KL equation is given by 
\begin{equation}    \label{eq:init_KL_eq}
    N_D(\rt,Y=Y_0,Y_0;\bt) = N^2(\rt,Y_0;\bt).
\end{equation}
Here $N(\rt,Y_0;\bt)$ is obtained as a solution to the BK equation.
The integral kernel in \cref{eq:KL_eq} is the one used in the BK equation (\ref{eq:BK_eq}) for $N(\rt,Y;\bt)$. The KL equation (\ref{eq:KL_eq}) for $N_D(\rt,Y,Y_0;\bt)$ can be transformed into the BK equation (\ref{eq:BK_eq}) for the quantity $N_I(\rt,Y,Y_0;\bt) \equiv 2N(\rt,Y;\bt) - N_D(\rt,Y,Y_0;\bt)$, which is the method we use to solve it numerically together with the BK evolution for $N(\rt,Y;\bt)$.

The diffractive cross section for the virtual photon-target scattering can be expressed in terms of the diffractive dipole-target cross section, similarly as in the inclusive case, as
 \begin{multline}    \label{eq:dip_factorization_diff}
     \frac{\dd {\sigma_{D\ (T,L)}^{\gamma^*h}}}{\dd{\ln (1/\beta)}} (\beta,\xpom,Q^2) = \sum\limits_f \int \dd[2]{\rt} \int\limits_0^1 \dd{z} \big\lvert \psi_{T,L}^{\gamma^*\to f\Bar{f}}\big\rvert^2 \\ \times \frac{\dd{\sigma^\mathrm{q\Bar{q}h}_D}}{\dd{\ln(1/\beta)}} (\beta,\xpom,{\bf r}).  
\end{multline}
The diffractive dipole-target cross section with a specific value of the gap is obtained as a derivative of $N_D$, which was defined as an integral over rapidity gap sizes greater than $Y_0$:
\begin{equation}
\label{eq:sigma_dip_diff_N_KL}
    \frac{\dd\sigma_{D}^\mathrm{q\Bar{q}h}}{\dd\ln(1/\beta)} = \int \dd[2]{\bf b}\ 
    \left. \left(-\frac{\dd N_D(\rt,Y,Y_0;\bt)}{\dd Y_0}\right)  \right|_{Y_{0} = \ygap}.
\end{equation}
The minus sign in the above formula is from the definition of $Y_0$ as the lower limit of possible gap sizes. Recall that the size of the rapidity gap at fixed Bjorken-$x$ is related to the mass of the diffractively produced system, see Fig.~\ref{fig:DDIS_illustration} and the definitions of the kinematic variables in Eqs.~\eqref{eq:defbeta} and~\eqref{eq:defxpom}.

The KL formulation provides an elegant way to analyse diffractive dissociation in the electron-hadron scattering at high-energy in the high-mass regime. We will hereafter treat the two cases (proton and nucleus) separately. We first apply the framework to proton targets. We then  generalize the dipole-proton amplitude to the dipole-nucleus case, following Ref.~\cite{Lappi:2013zma}, in  Sec.~\ref{sec:EIC-prediction}.

\section{Scattering off proton: comparison to HERA data}
\label{sec:hera_comparisons}

In deep inelastic scattering off a proton, we  assume that the impact parameter dependence completely factorizes from both $N$ and $N_D$, and only the $b$-independent parts are evolved by the BK and KL equations. A similar factorization is assumed in Refs.~\cite{Albacete:2010sy,Lappi:2013zma} 
 where the initial condition for the BK evolution of the dipole-proton amplitude is fitted to inclusive structure function data.  
Now the dipole amplitude can be written as $N(\rt,Y;\bt) = T_p(\bt)\mathcal{N}(r,Y)$, where $T_p(\bt)$ is a certain  transverse density profile and $\mathcal{N}(r,Y)$ satisfies the $\bt$-independent BK equation. 
After integrating over all impact parameters we obtain
\begin{equation}    \label{eq:norm_N}
    \int \dd[2]{\bf b} T_p(\bt) = \sigma_0/2.
\end{equation}
Here the effective transverse size of the proton is denoted by convention as $\sigma_0/2$ (to compensate the factor $2$ originating from the optical theorem in~\cref{eq:dipole_tot_cs}), and is constrained by the HERA structure function data together with the initial condition for the BK equation.

Similarly the impact parameter dependence of the diffractive cross section is assumed to factorize as
\begin{equation}
 \int \dd[2]{\bt} N_D(\rt,Y,Y_0;\bt)  =  \sigma^D_0 \mathcal{N}_D (r,Y,Y_0),
\end{equation}
where $\mathcal{N}_D (r,Y,Y_0)$ is independent of the impact parameter and obeys the KL equation, and $\sigma^D_0$ is a constant.
The normalization factor $\sigma_0^D$ can be deduced by noticing that at the initial condition of the KL evolution we have $N_D = N^2$, see Eq.~\eqref{eq:init_KL_eq}. This gives 
\begin{equation}
\label{eq:sigma0_d}
    \sigma^D_0 = \int \dd[2]{\bf b} T_p^2({\bf b}),
\end{equation}
and  implies that, for a given $\sigma_0$, $\sigma^D_0$ depends strongly on the shape of $T_p(\bt)$. Consequently the relative normalization of diffractive and inclusive cross sections depends on the assumed shape of the proton. 

The proton density profile can in principle be extracted from elastic scattering measurements. The spatial distribution of the small-$x$ gluon field is most directly probed in exclusive vector meson (e.g. $\mathrm{J}/\psi$) production measurements at HERA~\cite{Alexa:2013xxa,ZEUS:2004yeh}. This data is compatible with a Gaussian density profile $e^{-b^2/(2B)}$ with $B\approx 4\,\mathrm{GeV}^{-2}$, although a direct comparison is only possible with the factorized $b$-profile and becomes more involved if this approximation is relaxed~\cite{Demirci:2022wuy}. However, due to the limited squared momentum transfer $|t|$ region covered by these measurements, also other density profiles are possible, see e.g. Refs.~\cite{Frankfurt:2002ka,Mantysaari:2016jaz,Dumitru:2021hjm,Kowalski:2006hc,Kuokkanen:2011je}.

We parametrize the proton density profile using the regularized incomplete gamma function profile following Ref.~\cite{Rybczynski:2013mla}, with a the parameter $\omega$ controlling the steepness of the proton profile:  
\begin{equation}    \label{eq:b-profile-proton}
    T_p(\bt) = \frac{\Gamma\left(\frac{1}{\omega},\frac{b^2}{R_p^2\omega}\right)}{\Gamma\left(\frac{1}{\omega}\right)}.
\end{equation}
Here $\pi R_p^2 = \sigma_0/2$ and $w\ge 0$. At $\omega\to 0$, $T_p(b)|_{\omega \to 0} = \Theta(R_p-b)$ (hard sphere), while at $\omega = 1$ the profile becomes Gaussian, $T_p(b)|_{\omega = 1} = \exp\left(-b^2/R_p^2\right)$. The Gaussian form corresponds to the one usually employed in the literature, e.g. in the popular IPsat parametrization for the dipole-target scattering~\cite{Kowalski:2003hm}. The normalization factor for the  diffractive cross sections $\sigma_0^D$ defined in Eq.~\eqref{eq:sigma0_d} will vary around the corresponding value obtained in terms of a Gaussian profile,  $\sigma^D_0(\omega=1)=\sigma_0/4$, depending on how steep the profile is compared to the Gaussian shape. 

\begin{figure}[tb]
    \centering
    \includegraphics[width=\columnwidth,height=\textheight,keepaspectratio]{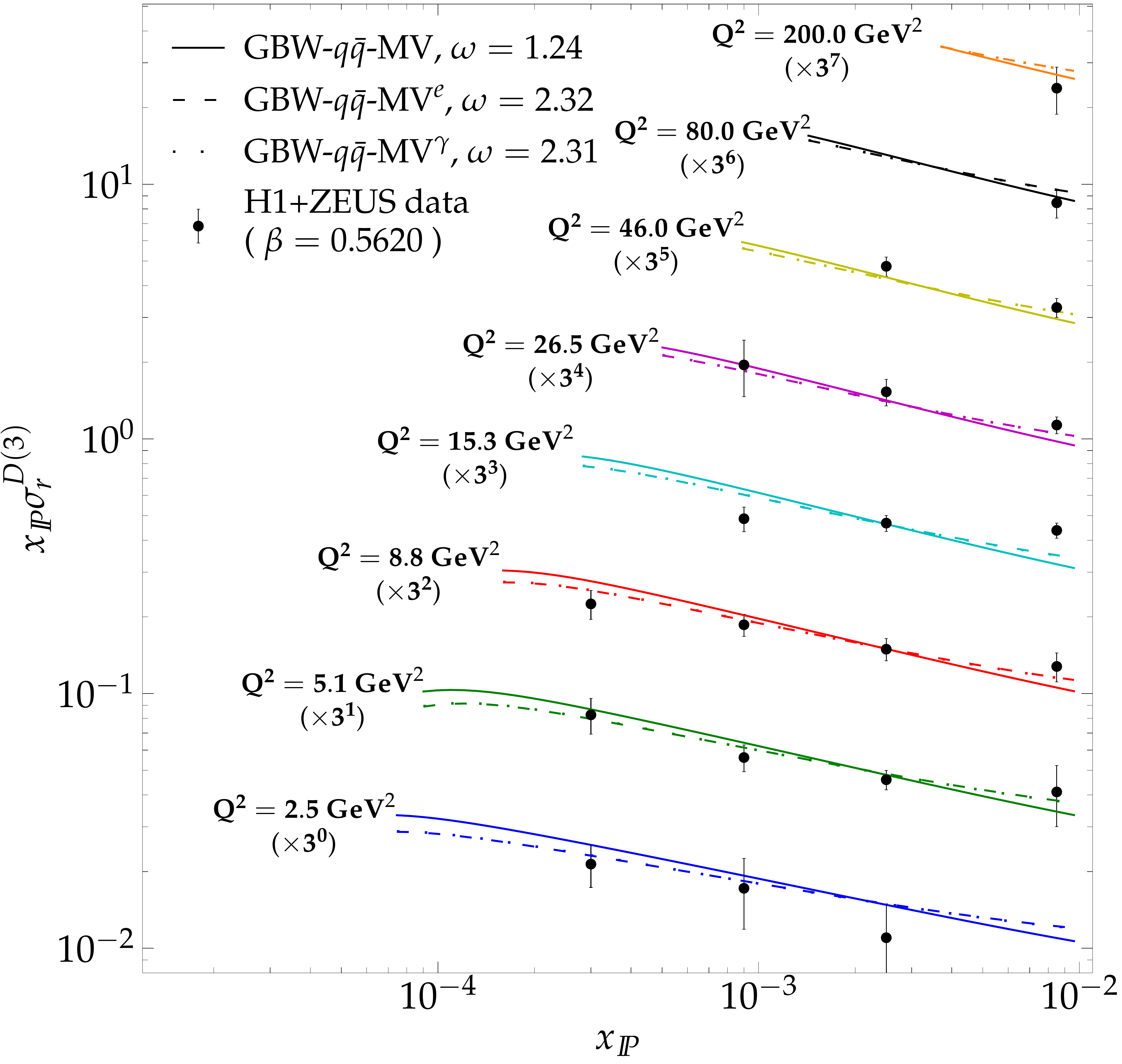}
    \caption{The reduced diffractive cross sections taking into account only the $q\Bar{q}$ contribution with the $\omega$ dependent normalization factor fitted to the HERA combined data~\cite{H1:2012xlc} at $\beta > 0.5$ and at different $Q^2$ bins. Only the results at $\beta=0.562$ are shown in this plot. The optimal values for $\omega$ obtained with different dipole-proton amplitudes are shown in the legend. }
    \label{fig:GBW_fit}
\end{figure}

\begin{figure}[tb]
    \centering
    \includegraphics[width=\columnwidth,height=\textheight,keepaspectratio]{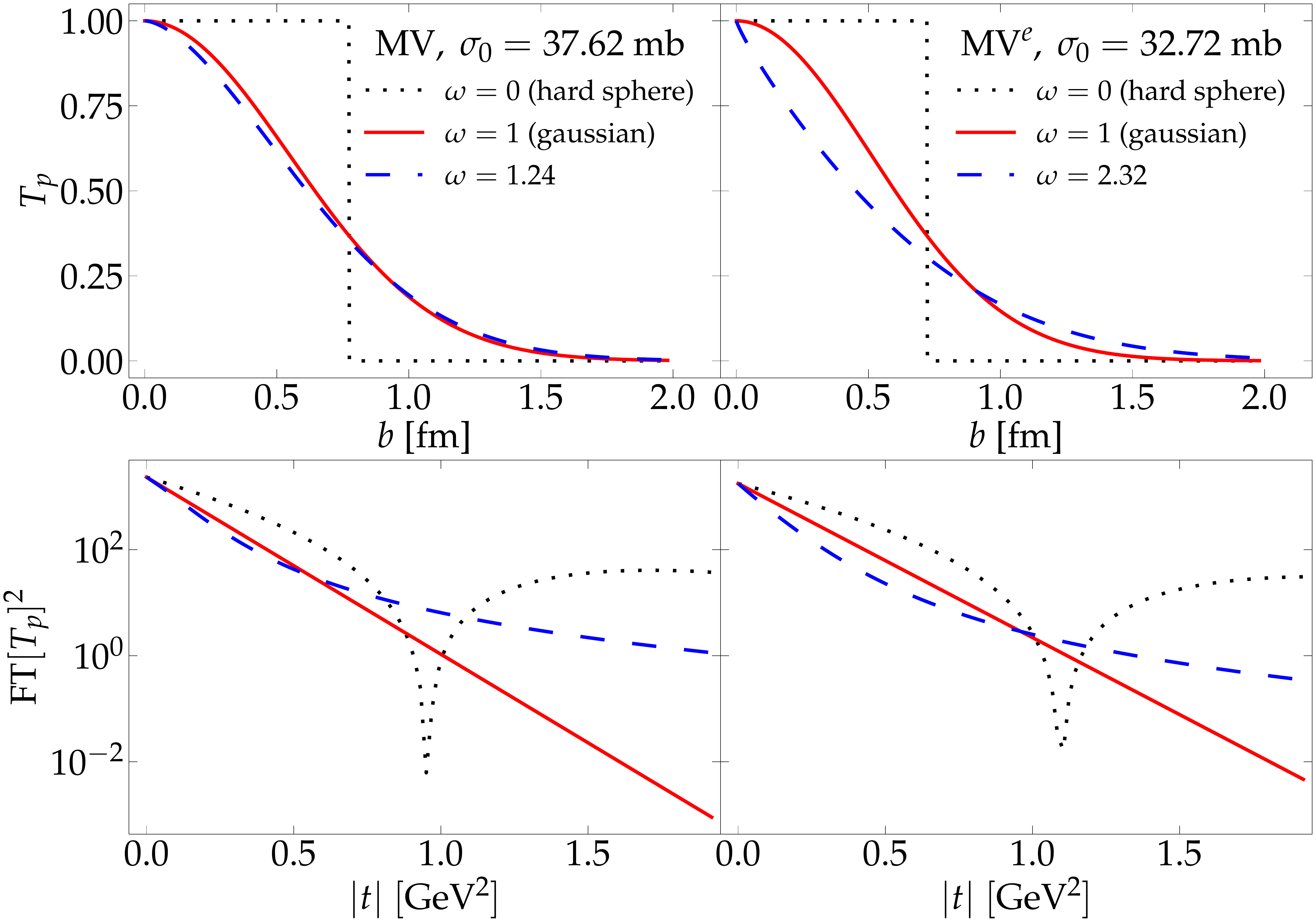}
    \caption{The proton impact parameter profiles given in~\cref{eq:b-profile-proton} for the determined optimal values of $\omega$ and $\sigma_0$. Their corresponding squared Fourier transforms (FT) are plotted in the second row.
    }
    \label{fig:proton_profile}
\end{figure}

\begin{figure*}[ht!]
    \centering
    \includegraphics[width=\textwidth,height=\textheight,keepaspectratio]{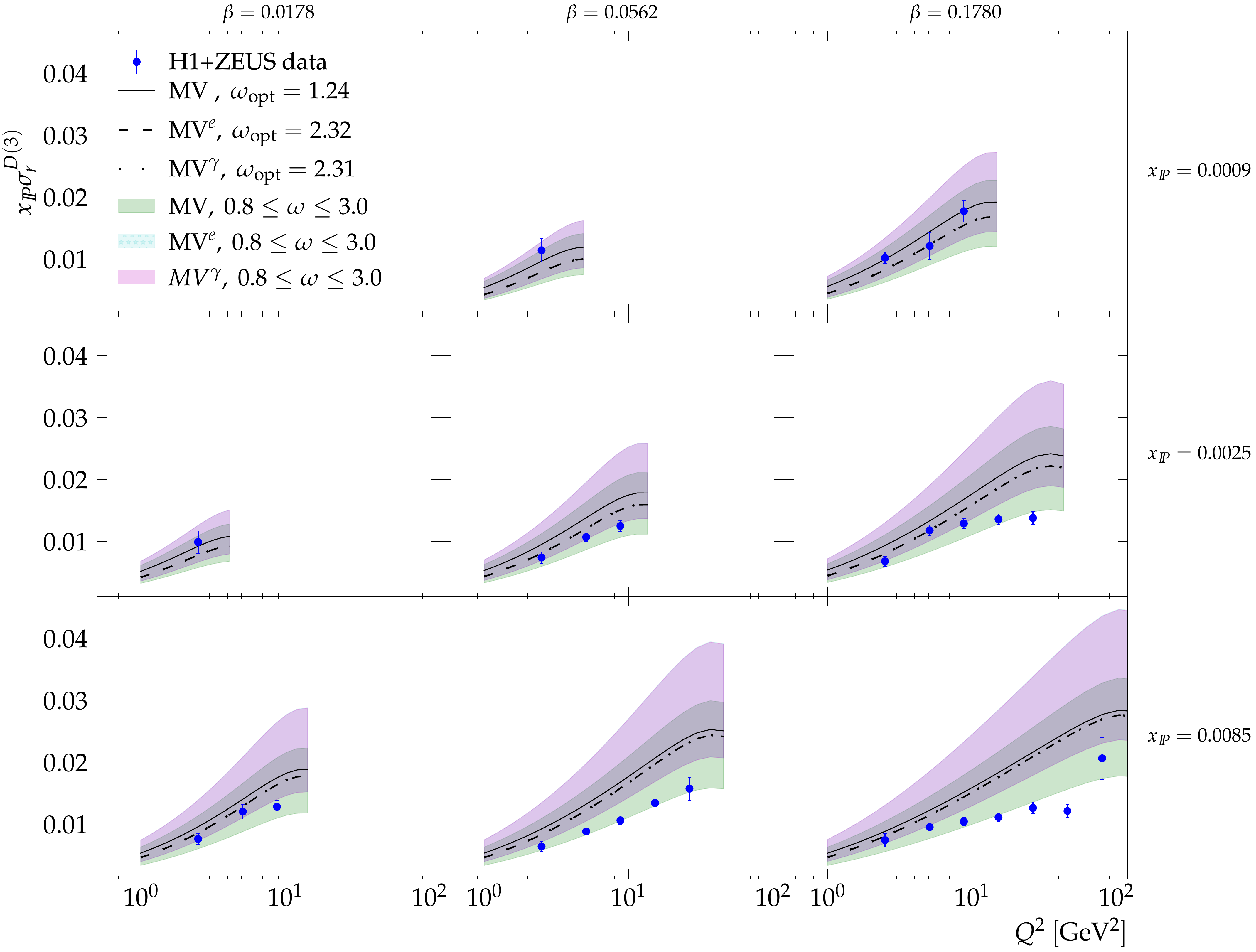}
    \caption{Diffractive reduced cross section as a function of $Q^2$ at different values of $\beta$ and $\xpom$. The HERA combined dataset is taken from Ref.~\cite{H1:2012xlc}. The bands represent the results from the KL solutions with the corresponding initial conditions for $\omega$ varying in the region $0.8\le \omega < 3$. The lines represent the numerical results for the optimal values of $\omega$ as explained in the text.}
    \label{fig:sigma_red_Q2}
\end{figure*}

\begin{table*}[ht!] 
        \centering
        \begin{tabular}{l | c c c c c | c} 
        \hline
        Parametrization & $Q_{s0}^2 (\rm GeV^2)$ & $\gamma$ & $e_c$ & $\sigma_0/2\ ({\rm mb})$ & $C^2$ & $\ \omega_\mathrm{opt}$ \\ [0.5ex] 
        \hline\hline
        MV & 0.104 & 1 & 1 & 18.81 & 14.5 &\ 1.24\\ 
        MV$^e$ & 0.060 & 1 & 18.9 & 16.36 & 7.2 &\ 2.32 \\ 
        MV$^\gamma$ & 0.159 & 1.129 & 1 & 16.35 & 7.05 &\ 2.31 \\[1ex]
        \hline
        \end{tabular}
        \caption{Parameters for the dipole-proton scattering amplitude \eqref{eq:mvparam} at the initial condition for the BK evolution used in the calculation (from Refs.~\cite{Lappi:2013zma,Albacete:2010sy}). The determined optimal values for the parameter $\omega$ in Eq.~\eqref{eq:b-profile-proton} controlling the shape of the proton density profile are also shown. }
        \label{tab:params}
\end{table*}

As mentioned above, the BK evolution starts with an initial amplitude at initial evolution rapidity $Y=Y_\mathrm{min}$ corresponding to $x=x_\mathrm{init} = 0.01$, at which we shall employ the following parametrization~\cite{Lappi:2013zma} based on the  McLerran-Venugopalan (MV) model~\cite{McLerran:1993ni}:
\begin{equation}
\label{eq:mvparam}
    \mathcal{N}(r) = 1 - \exp\left[ -\frac{(r^2Q_{s0}^2)^\gamma}{4}\ln\left(e\cdot e_c + \frac{1}{r\lqcd}\right)\right].
\end{equation}
Here $Q_{s0}^2$ controls the  initial proton saturation scale, $\gamma$ is the initial anomalous dimension, and $e_c$ modifies the behavior at large $r$. Their values used in this analysis are taken from the fits to the HERA inclusive structure function data~\cite{H1:2009pze} reported in Ref.~\cite{Lappi:2013zma} (see also the earlier similar study in Ref.~\cite{Albacete:2010sy}) and are summarized in \cref{tab:params}.
In addition, the constant $C^2$ controlling the scale of the coordinate space running coupling in Eq.~\eqref{eq:running_coupling} and the effective proton area $\sigma_0/2$ are also obtained from the corresponding fits. 
In this work we use all these three fits in order to determine the potential sensitivity on the uncertainties in the dipole-proton scattering amplitude.

For the current analysis, we consider the ZEUS FPC~\cite{ZEUS:2005vwg,ZEUS:2008qxs} and the H1 + ZEUS combined datasets~\cite{H1:2012xlc} for the diffractive structure functions and reduced cross sections. The combined data corresponds to coherent diffraction, as does our calculation. We use it to determine the optimal value for the proton shape parameter $\omega$ denoted by $\omega_\mathrm{opt}$. The ZEUS FPC data on the other hand contains a contribution from events where the proton dissociates to a system with relatively small invariant mass. When comparing to the ZEUS FPC data we scale the data down by a factor of $1.88$ following a heuristic procedure to be specified later in order to obtain an estimate for the coherent contribution.

The optimal proton shape parameter $\omega_\mathrm{opt}$ is determined as follows. We use the GBW result, Eqs.~\eqref{eq:gbw_qq_T} and~\eqref{eq:gbw_qq_L}, to calculate the diffractive cross section at high $\beta$ where the considered $q\Bar{q}$ component dominates~\cite{Kowalski:2008sa}. 
The optimal $\omega_\mathrm{opt}$ is then obtained by minimizing $\chi^2$ to the high-$\beta$ combined HERA data. 
We do not include the $q\Bar{q}g$ component here, as it gives a negligible contribution at high $\beta$, and  there is also an ambiguity in the scale of the running coupling. By fitting to the reduced diffractive cross section data at $\beta > 0.5$ ($24$ data points with $\beta = 0.562$ and $\beta = 0.816$, note that we only include the points with $\xpom \le 0.01$), we obtain $\omega_\mathrm{opt} \simeq 1.24\ (\chi_\mathrm{red}^2 \approx 1.87)$ for the MV, $\omega_\mathrm{opt} \simeq 2.32\ (\chi_\mathrm{red}^2 \approx 1.08)$ for the MV$^{e}$, and $\omega_{opt} \simeq 2.31\ (\chi_\mathrm{red}^2 \approx 1.09)$ for the MV$^{\gamma}$ parametrizations for the dipole-proton amplitude. Here $\chi_\mathrm{red}^2$ is $\chi^2$ per degree of freedom. The obtained good agreement with the $\beta=0.562$ data is shown in Fig.~\ref{fig:GBW_fit}. The modified MV model parametrizations MV$^e$ and MV$^\gamma$ result in almost identical cross sections and values for the proton shape  parameter $\omega\approx 2.3$ which is much steeper than the corresponding density profile with $\omega\approx 1.2$  obtained using the MV model fit. 

The density profiles corresponding to the optimal values of the $\omega$ parameter compared to the Gaussian and step function profiles are shown in \cref{fig:proton_profile}. In coordinate space the profile obtained with the MV model parametrization ($\omega=1.24$) is very close to a Gaussian one, and with $\omega=2.32$ corresponding to MV$^e$ and MV$^\gamma$ fits for the dipole amplitude we obtain a density profile that is much more steeply falling than a Gaussian close to the center of the proton, but which has a longer large-$b$ tail. The corresponding two dimensional Fourier transforms are also shown in \cref{fig:proton_profile} as a function of $t=-\Deltat^2$, where $\Deltat$ is the Fourier conjugate to the impact parameter. Note that the exclusive vector meson production cross section discussed above is approximatively proportional to the squared Fourier transform.  In  Fourier space the $\omega=1.24$ and the Gaussian profiles only deviate significantly in the $|t|\gtrsim 0.5\gev^2$ region where there is only limited data available, while for $\omega=2.32$ the $|t|$-spectrum is somewhat steeper. We also note that with $\omega>1$ we do not obtain any diffractive dips, and recall that no such minima are visible in the HERA data up to $|t|\sim 1\gev^2$. 
For a detailed discussion about the diffractive minima and their potential relation to saturation effects, see also Ref.~\cite{Armesto:2014sma}.

\begin{figure*}[ht!]
    \centering
    \includegraphics[width=\textwidth,height=\textheight,keepaspectratio]{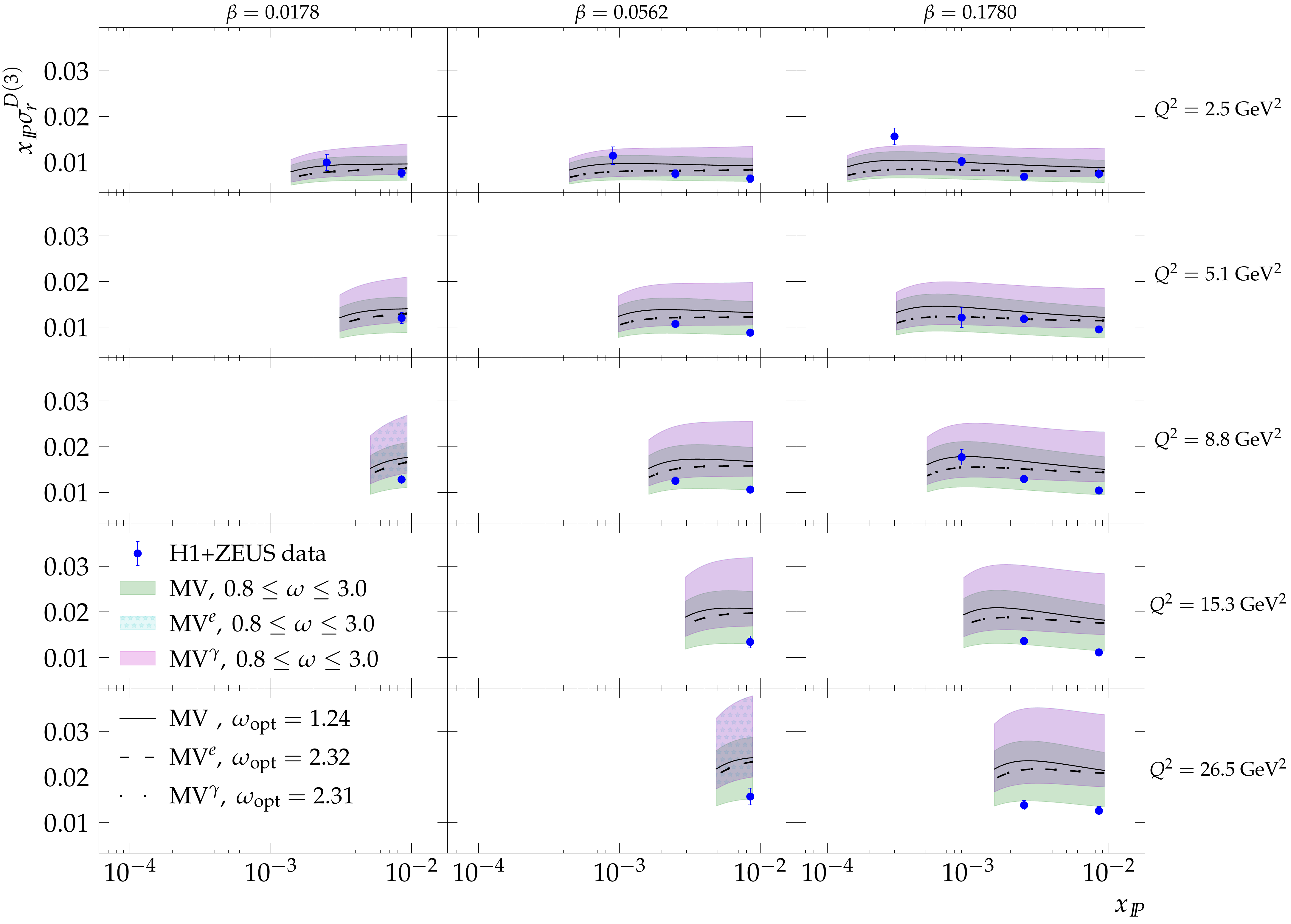}
    \caption{Diffractive reduced cross section as a function of $\xpom$ at different values of $\beta$ and $Q^2$. The HERA combined dataset is taken from Ref.~\cite{H1:2012xlc}. The notations are same to~\cref{fig:sigma_red_Q2}. }
    \label{fig:sigma_red_xP}
\end{figure*}

Next we use the determined proton density profiles and compute predictions for the diffractive reduced cross section in a wide kinematical domain covered by the combined HERA data~\cite{H1:2012xlc}, now using the result obtained by solving the Kovchegov-Levin equation as disucssed in \cref{sec:kl}.  
The reduced cross section as a function of $Q^2$ in different bins of $\xpom$ and $\beta$ is shown  in~\cref{fig:sigma_red_Q2}. The KL solutions exhibit a visible rise in $Q^2$, for all values of $\beta$ and $\xpom$, up to a large  $Q^2$ where $y\gtrsim 0.5$ and the second term in~\cref{eq:red_cross_section} becomes dominant. At $\beta>0.1$, the data however depend weakly on $Q^2$, which agrees with the known leading-twist behavior of the quark-antiquark contribution. The KL solutions cannot describe appropriately the data in this region.
At smaller $\beta$, where the effect of (soft) gluon emissions becomes important,
a better description of the combined HERA data is obtained using the KL perturbative evolution equation, although the cross section especially at higher $\xpom$ is typically slightly overestimated. 

The dependence on the proton shape parameter is also illustrated in \cref{fig:sigma_red_Q2} (and the figures following)  by varying the $\omega$ parameter around the optimal value. 
Similarly to the large-$\beta$ case, the normalization of the diffractive cross section is typically well described with $\omega > 1$, and as such also the small-$\beta$ data prefers a density profile which is steeper than Gaussian, corresponding to a smaller overall normalization for the diffractive cross section.

The reduced diffractive cross section as a function of $\xpom$ is shown in Fig.~\ref{fig:sigma_red_xP}. Again a good agreement with the data is obtained at (moderately) small $\beta$, although the normalization at high $Q^2$ is typically overestimated as already seen above in \cref{fig:sigma_red_Q2}.
The maximum in the reduced cross section observed at small $\xpom$ is again due to the longitudinal cross section $F_L^{D(3)}$ becoming important when $y \gtrsim 0.5$. 
The $\xpom$ dependence becomes milder toward smaller $\beta$ and smaller $Q^2$. The mild $\xpom$ dependence seen especially at small virtualities is  compatible with the predictions from the BK and KL equations.  

\begin{figure}[tb]
    \centering
    \includegraphics[width=\columnwidth,keepaspectratio]{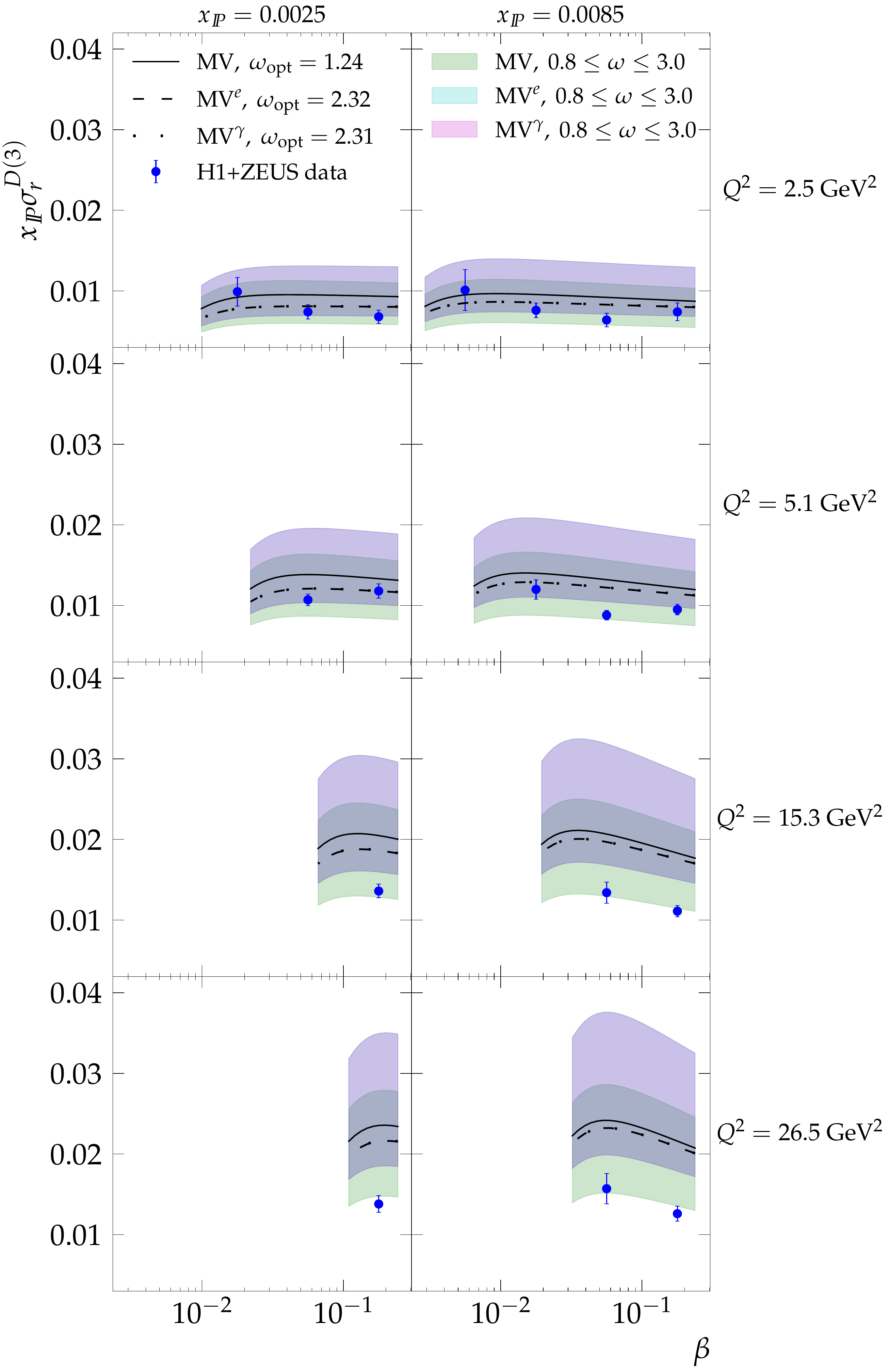}
    \caption{Diffractive reduced cross section as a function of $\beta$ at different values of $\xpom$ and $Q^2$. The HERA combined dataset is taken from Ref.~\cite{H1:2012xlc}. The notations are the same as in \cref{fig:sigma_red_Q2}. 
    }
    \label{fig:sigma_red_beta}
\end{figure}

\begin{figure}[tb]
    \centering
    \includegraphics[width=\columnwidth,height=\textheight,keepaspectratio]{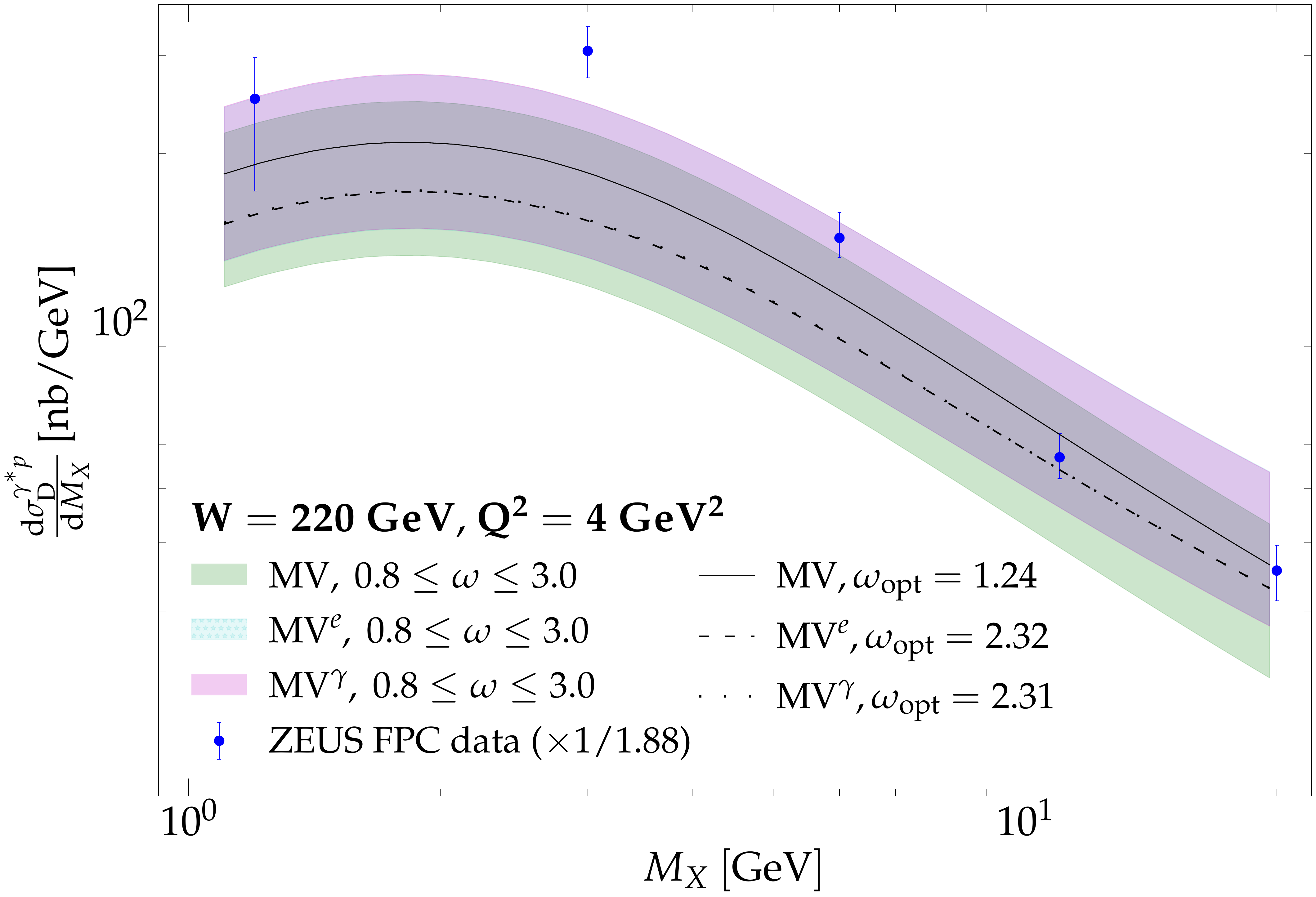}
    \caption{Mass spectrum at $W=220$GeV and $Q^2=4$GeV$^2$ compared to the ZEUS FPC data~\cite{ZEUS:2005vwg} (scaled down by a factor of 1.88). The bands are the results from the solutions to the KL equations with the corresponding initial conditions, and with the b-profile parameter $\omega$ varying in the range $0.5 \le \omega \le 2.0$. The lines represent the numerical results for the optimal values of $\omega$ as explained in the text. }
    \label{fig:dsigma_MX_proton}
\end{figure}

To directly probe the $\ln 1/\beta$ evolution described by the KL equation we also calculate the diffractive cross section as a function of the mass of the diffractively produced system $M_X$ or $\beta$ (recall that $M_X^2/Q^2\sim 1/\beta$).
The results as a function of $\beta$ compared with the combined HERA data are shown in \cref{fig:sigma_red_beta}, and as a function of $M_X$ compared with the  
ZEUS FPC dataset~\cite{ZEUS:2005vwg,ZEUS:2008qxs}  in \cref{fig:dsigma_MX_proton}. 
As mentioned before, the ZEUS FPC data includes some contribution from incoherent events where the proton dissociates into a low mass state ($\gamma^*+p\to X+N$, $M_N<2.3\ \rm GeV$). In order to approximatively remove this dissociative contribution not included in our calculation we scale down the data by a constant factor of $1.88$. This factor is obtained as follows. First, the original ZEUS FPC data with $\beta>0.5$ ($154$ points) are fitted using the GBW result with only the $q\Bar{q}$ contribution to obtain the optimal value for $\omega$ for each initial condition. We then compute the ratio between $\sigma_0^D$ at the obtained $\omega$ and the one at $\omega_{\rm opt}$ obtained from the fit to the HERA combined data above. 
The three different fits for the initial conditions of the BK evolution result in very similar ratio, and the average value 
$1.88$ is then chosen to be the scaling factor\footnote{We note that a slightly smaller value has been used in previous analyses e.g. in Ref.~\cite{Kowalski:2008sa}.} 

Again we find a good description of the available data, although the cross section is typically overestimated at high $Q^2$. More importantly the $\beta$ and $M_X$ dependencies predicted by the KL equation are compatible with the HERA data, when we focus on the moderately high-mass regime ($\beta \lesssim 0.1$). 

The mass spectra at fixed $W$ and $Q^2$ from the numerical calculation shown in \cref{fig:dsigma_MX_proton} exhibit a similar trend as the data, which decreases toward the high-mass (small $\beta$) regime at a fixed Bjorken $\xbj$. Given the very mild dependence of the diffractive struction function on $\beta$ as shown above, this behavior is predominantly due to the $M_X$-dependent prefactor in~\cref{eq:mass_spectrum_def}. Up to the chosen scaling factor, the KL evolution describes the mass dependence well in the high-mass domain. The diffractive cross section is underestimated in the low-mass domain, but we again emphasize that the KL evolution is expected to be an accurate description of the QCD dynamics only in the high-$M_X$ region. However, a qualitative description of the data is also obtained when the KL results are extrapolated to the low-$M_X$ region. 

\begin{figure}[tbp!]
    \centering
    \includegraphics[width=\columnwidth,height=\textheight,keepaspectratio]{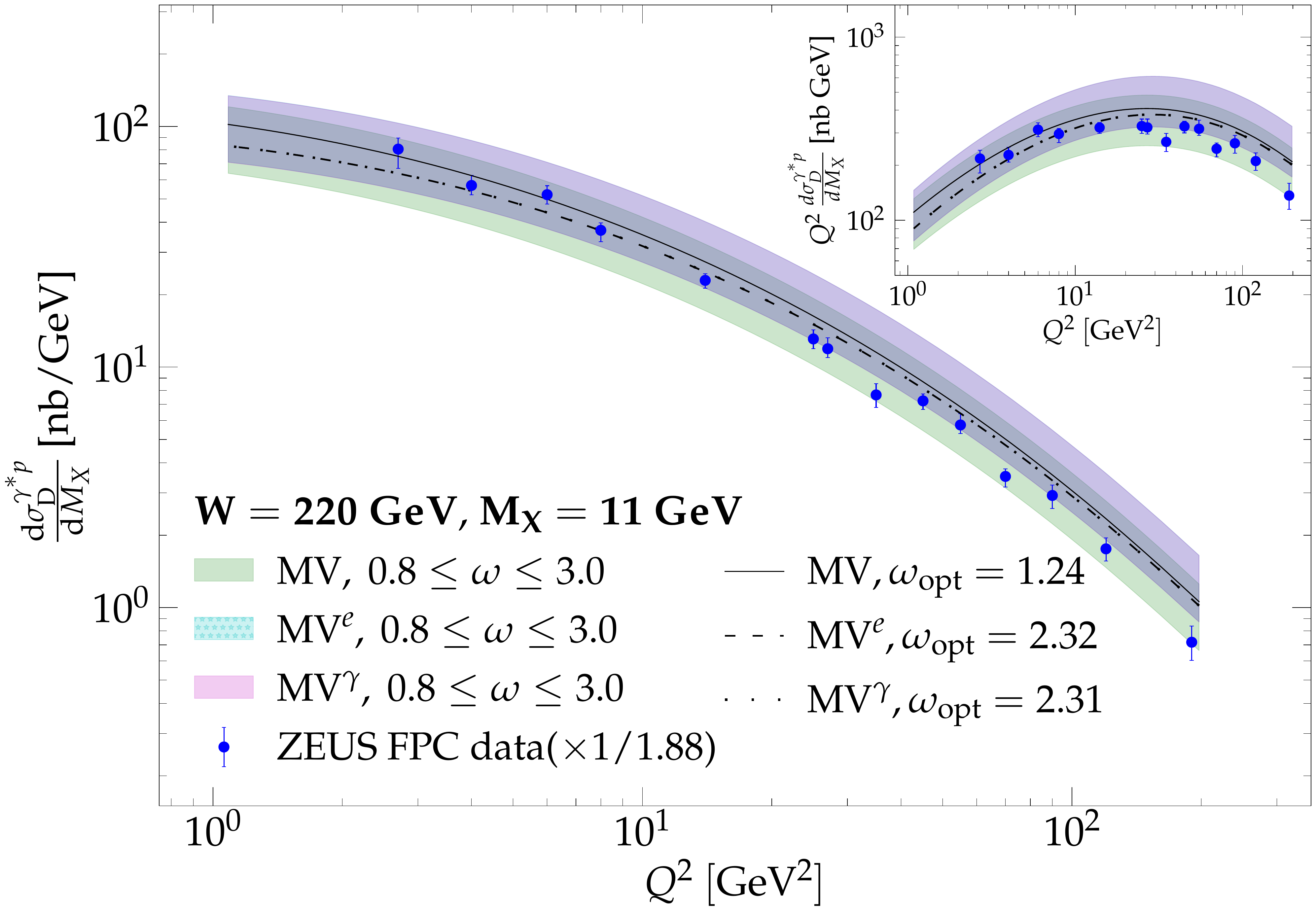}
    \caption{The dependence of the diffractive cross section  on $Q^2$  at $W=220$GeV and $M_X=11$GeV compared to the ZEUS FPC data~\cite{ZEUS:2005vwg,ZEUS:2008qxs}  (scaled down by a factor of 1.88). The inset shows the diffractive cross section scaled by the photon virtuality, $Q^2\dd\sigma^{\gamma^*p}/\dd{M_X}$. }
    \label{fig:dsigma_Q2_proton}
\end{figure}

\begin{figure*}[tbp!]
    \centering
    \includegraphics[width=\textwidth,height=\textheight,keepaspectratio]{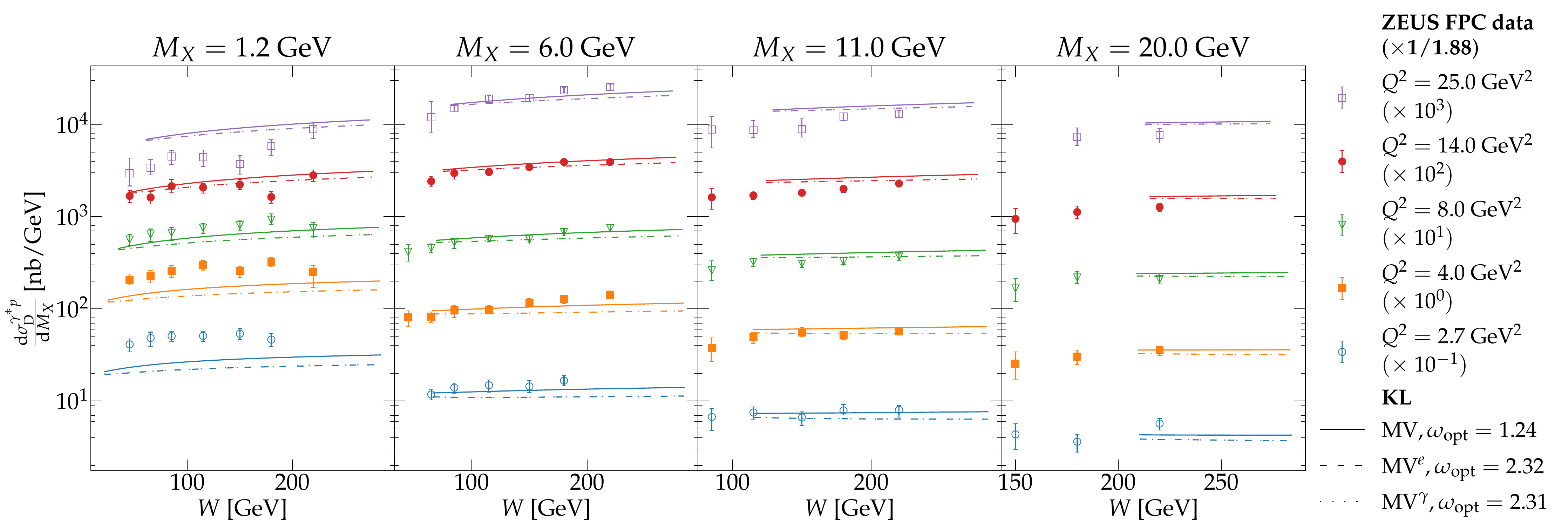}
    \caption{$W$ dependence of the diffractive cross section at different diffractive masses $M_X$ and photon virtualities $Q^2$ compared to the ZEUS FPC data~\cite{ZEUS:2005vwg} (scaled down by a factor of 1.88). For simplicity, we show only the results with obtained with the optimal values of $\omega$. }
    \label{fig:dsigma_W}
\end{figure*}

To complete our comparisons with the available HERA data, let us finally compare the $Q^2$ and $W$ dependencies obtained from the solutions to the KL and BK equations to the ZEUS FPC data. 
The virtuality dependence at relatively high $M_X$ is shown in \cref{fig:dsigma_Q2_proton}, and the center-of-mass-energy $W$ dependency is shown in \cref{fig:dsigma_W}. 
Similarly as when comparing to the combined HERA data, the $Q^2$ and $W$ dependencies in the ZEUS data are described fairly well especially when  $Q^2$ is not very large (i.e., $\beta$ is small). 
While the cross section changes mildly with $W$ in general, there is a significant decrease with increasing $Q^2$.
Such a  decrease together with a modest variation of the scaled diffractive cross-section, $Q^2\dd\sigma^{\gamma^*p}_D/\dd M_X$ (shown in the inset of~\cref{fig:dsigma_Q2_proton}), for $Q^2<M_X^2$ are  indications for a leading twist-like behavior.

\begin{figure*}[th!]
    \centering
    \includegraphics[width=\textwidth,height=\textheight,keepaspectratio]{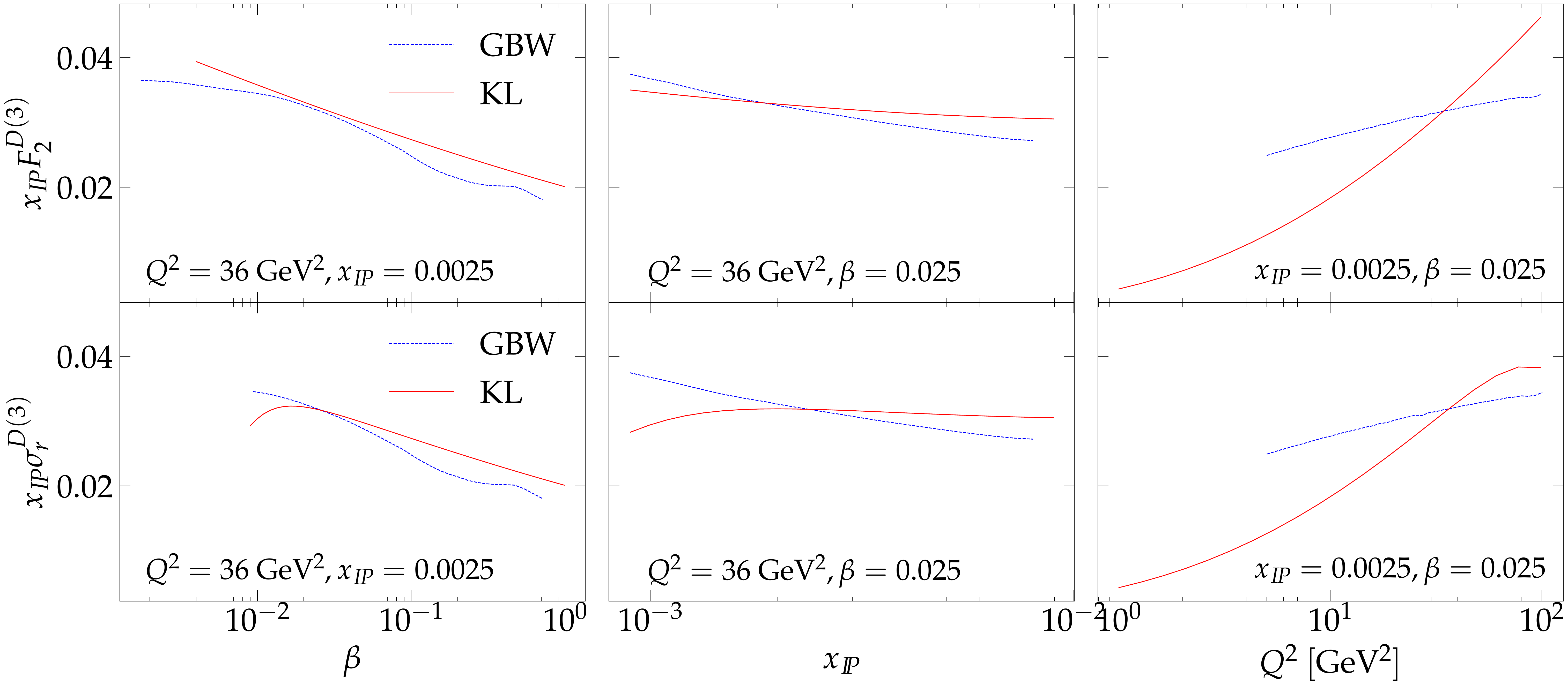}
    \caption{Comparison of the KL and the GBW results for diffractive structure function (first row) and the diffractive reduced cross section (second row) at $\sqrt{s}=1.3\,\mathrm{TeV}$ using the ${\rm MV}^{e}$ initial condition. Three columns show the dependences on $\beta$, $\xpom$ and $Q^2$, respectively. }    
    \label{fig:KL_vs_GBW}
\end{figure*}

Before ending this section, let us compare the KL calculation to the GBW results including both the $q\Bar{q}$ and $q\Bar{q}g$ contributions.
We emphasize that these results are strictly speaking valid in different kinematical limits: the GBW result including the $q\Bar{q}g$ contribution given by \cref{eq:gbw_qqg} is valid at high-$Q^2$ and the KL evolution dominates at low-$\beta$. 
The calculations are performed in the kinematics with $\sqrt{s} = 1.3\ \rm TeV$, which could be accessible in the future experiments such as the LHeC/FCC-eh, in order to have a wider phase space available.
The comparison is shown in~\cref{fig:KL_vs_GBW}. The diffractive structure function scaled by $\xpom$ rises toward small $\xpom$, small $\beta$ and large $Q^2$ in both approaches. As for the diffractive reduced cross sections, there is however a peak in the region with $y\gtrsim 0.5$ for the KL solutions, which does not manifest in the GBW result. This is attributed to the fact that the longitudinal contribution from gluon-dressed states is not included in the latter.  

The $\beta$ dependence from the GBW and the KL approaches is similar in the moderately small $\beta$ region. The large-$\beta$ structure  in the GBW results originates from the different components ($q\bar q$ from longitudinal or transverse photon, or $q\bar q g$) dominating at different $\beta$ values~\cite{Kowalski:2008sa}. At very small $\beta \lesssim 10^{-2}$ the higher Fock states resummed in the KL evolution become important and result in faster increase of the cross section with decreasing $\beta$ compared to the GBW approach.

The more obvious differences between the two results can be seen in the $\xpom$ and $Q^2$ spectra. To understand these discrepancies, let us return the formalism of the two approaches. The KL evolution is basically a BK evolution with a small delay at $Y_0$. This delay will not change the dominant shape of the BK front in the dilute regime, meaning that the solutions to the KL in such regime scale as $\mathcal{N}_D(\rt,Y,Y_0) \sim \left[\rt^2Q_{s,D}^2(Y,Y_0)\right]^{\gamma_c}$ as for the BK, where $\gamma_c\approx 0.85$ is the anomalous dimension generated by the running-coupling BK evolution~\cite{Albacete:2007yr}.
Here $Q_{s,D}^2$ refers to the saturation scale extracted from the diffractive cross section obtained as a solution to the KL equation. 
Note that here the delay does modify the saturation scale, which turns out to be its main effect, so that the saturation scale now depends very mildly on $Y_0$, as shown numerically in Ref.~\cite{Levin:2001pr}. Convoluting with the squared photon wave functions (see Ref.~\cite{Hatta:2006hs} for the detailed treatment of the $\br$-integration) and considering $Q^2>Q_{s,D}^2$ (which is relevant to our analyses), the diffractive structure function behaves as
\begin{equation}
\label{eq:f2d_kl_behavior}
    \left[F_{2}^{D(3)}\right]_\mathrm{KL} \sim Q^2 \left(\frac{Q_{s,D}^2}{Q^2}\right)^{\gamma_c},  
\end{equation}
with the extra $Q^2$ from~\cref{eq:f2D_def}. In this case, the dominant contribution to the $\br$-integration comes from the dipole sizes $r\sim 1/Q$. Again, \cref{eq:f2d_kl_behavior} can explain the $Q^2$ behavior of $\dd\sigma^{\gamma^*p}_D/\dd M_X$ (without the extra $Q^2$) shown in \cref{fig:dsigma_Q2_proton}.

Now we turn to the GBW result. Taking the $q\Bar{q}$ contribution, the diffractive cross section scales as the dipole-proton amplitude squared $\mathcal{N}^2(\rt,Y) \sim \left[\rt^2Q_{s}^2(\xpom)\right]^{2\gamma_c}$, with $Q_{s}$ now being the normal saturation momentum from the BK evolution evaluated at $\xpom$. The $\br$-integration leads to
\begin{equation}
\label{eq:f2d_gbw-qq_behavior}
    \left[F_{q\Bar{q}}^{D(3)}\right]_\mathrm{GBW} \sim Q^2 \left(\frac{Q_{s}^2}{Q^2}\right) = Q_{s}^2.  
\end{equation}
Meanwhile, the contribution of the $q\bar{q}g$ component is given by \cite{Munier:2003zb,Hatta:2006hs}
\begin{equation}
\label{eq:f2d_gbw-qqg_behavior}
   \left[F_{q\Bar{q}g}^{D(3)}\right]_\mathrm{GBW} \sim Q^2 \left(\frac{Q_{s}^2}{Q^2}\right) \ln \frac{Q^2}{Q_{s}^2} = Q_{s}^2\ln \frac{Q^2}{Q_{s}^2}. 
\end{equation}
Unlike the KL case, the $\rt$-integration leading to \cref{eq:f2d_gbw-qq_behavior,eq:f2d_gbw-qqg_behavior} is dominated by $r\sim 1/Q_s$.

Some remarks are in order concerning \cref{eq:f2d_kl_behavior,eq:f2d_gbw-qq_behavior,eq:f2d_gbw-qqg_behavior}. First, the diffractive structure function from the KL evolution has a power-law behavior in $Q^2$, which grows faster than the logarithmic shape of the same observable calculated from the GBW approach. Furthermore, the KL evolution results in a milder dependence on $\xpom$ of the diffractive structure function compared to the GBW calculation. Such behaviors can be indeed observed in the numerical comparison shown in~\cref{fig:KL_vs_GBW}. Finally, it is interesting to note that the further additions of gluons to the dipole wave function make the $Q^2$-dependence become steeper, which manifests itself in the the transition between the two approaches when varying $\beta$.

To conclude this comparison, we note that the resummation of soft gluons included in the KL evolution has a significant effect on the $\beta$ dependence of the cross section only in the very small $\beta \lesssim 10^{-2}$ region which is only accessible in very high-energy nuclear DIS experiments such as the LHeC/FCC-he. On the other hand, the KL evolution also has a significant effect on the $\xpom$ and $Q^2$ systematics already in the EIC energy range, and as such the future EIC measurements will be able to (at least indirectly) probe the KL evolution dynamics.

\section{Electron-nucleus scattering:  predictions for the future EIC}
\label{sec:EIC-prediction}

Now let us move from a proton to a nuclear target.
Unlike in the proton case, we do not assume that the impact parameter dependence factorizes from the dipole-nucleus scattering amplitude. However, instead of investigating the fully impact-parameter-dependent BK and KL evolution equations, we follow Ref.~\cite{Lappi:2013zma} and solve these equations at each impact parameter $b=|\bt|$ independently. 
This approximation both simplifies the numerical calculation and also automatically avoids the problem of unphysical Coulomb tails which need to be regularized if finite-size effects are included in the evolution~\cite{Berger:2010sh,Mantysaari:2018zdd,Schlichting:2014ipa}.

The initial condition for the BK evolution of the dipole-nucleus amplitude at fixed impact parameter 
 is obtained by generalizing the dipole-proton scattering amplitude using the optical Glauber model following Ref.~\cite{Lappi:2013zma} to obtain
\begin{multline}    \label{eq:init_BK_A}
    N_A(r,b) = 1 - \exp\left[-AT_A(b)\frac{\sigma_0}{2}\frac{(r^2Q_{s0}^2)^{\gamma}}{4} \right.\\
    \left. \times \ln\left(e\cdot e_c + \frac{1}{r\lqcd}\right)\right].
\end{multline}
Here the subscript ``A" is used to distinguish with the same quantities in the proton case. The nuclear thickness function $T_A(b)$ is obtained from the Wood-Saxon (WS) distribution
\begin{equation}
    \rho_A({\bf b},z) = \frac{\rho_0}{1+\exp\left[\frac{\sqrt{\bt^2+z^2}-R_A}{d}\right]}  
\end{equation}
by integrating over the longitudinal coordinate $z$. The nuclear geometry is controlled by the parameters  $d=0.54\ \rm fm$ and $R_A= (1.12A^{1/3} - 0.86A^{-1/3})\ \rm fm$, and $\rho_0$ is obtained from the normalization condition $\int \dd[2]{\bf b} T_A(|{\bf b}|) = 1$. As discussed in Ref.~\cite{Lappi:2013zma}, this approach results in nuclear effects vanishing for small dipoles at the initial condition of the BK evolution.
The other parameters in \cref{eq:init_BK_A} are from the fits to the inclusive HERA data discussed in \cref{sec:hera_comparisons}. We will hereafter denote these by Glauber-${\rm MV}$, Glauber-${\rm MV}^e$ and Glauber-${\rm MV}^{\gamma}$  initial conditions originating from the ${\rm MV}$, ${\rm MV}^e$ and ${\rm MV}^{\gamma}$ proton fits, respectively. 

Following Ref.~\cite{Lappi:2013zma} we note that the nuclear saturation scales fall below the proton saturation scales at $b \gtrsim 6.45 \ \rm fm$ (Glauber-MV) and $b \gtrsim 6.3\ \rm fm$ (Glauber-MV$^e$ and Glauber-MV$^\gamma$). The BK evolution would result in a gluon density increasing rapidly in this low density region, which would lead to unphysically rapid growth of the nuclear size. 
Consequently in this dilute regime ($b>b_\mathrm{cut}$) we do not use the solutions to the evolution equations for the nuclear target, but assume that the nuclear scattering is an incoherent sum of the scatterings off nucleons which is also known as the impulse approximation (IA). This gives 
\begin{equation}
    N_A(r,Y;b>b_\mathrm{cut}) =  A T_A(b) \frac{\sigma_0}{2}\mathcal{N}(r,Y) .
\end{equation}
The scaling of the diffractive dipole-nucleus cross section $N_{D,A}$ (see \cref{eq:sigma_dip_diff_N_KL}) in this regime can be deduced from the initial condition  of the KL equation, \cref{eq:init_KL_eq}, and reads
\begin{equation}
\label{eq:dipole_nucleus_diffractive_largeb}
    N_{D,A}(r,Y,Y_0;b>b_{\rm cut}) = A^2T_A^2(b)\frac{\sigma_0^2\mathcal{N}_D(r,Y,Y_0)}{4}.
\end{equation}

\begin{figure*}[ht!]
    \centering
    \includegraphics[width=\textwidth,height=\textheight,keepaspectratio]{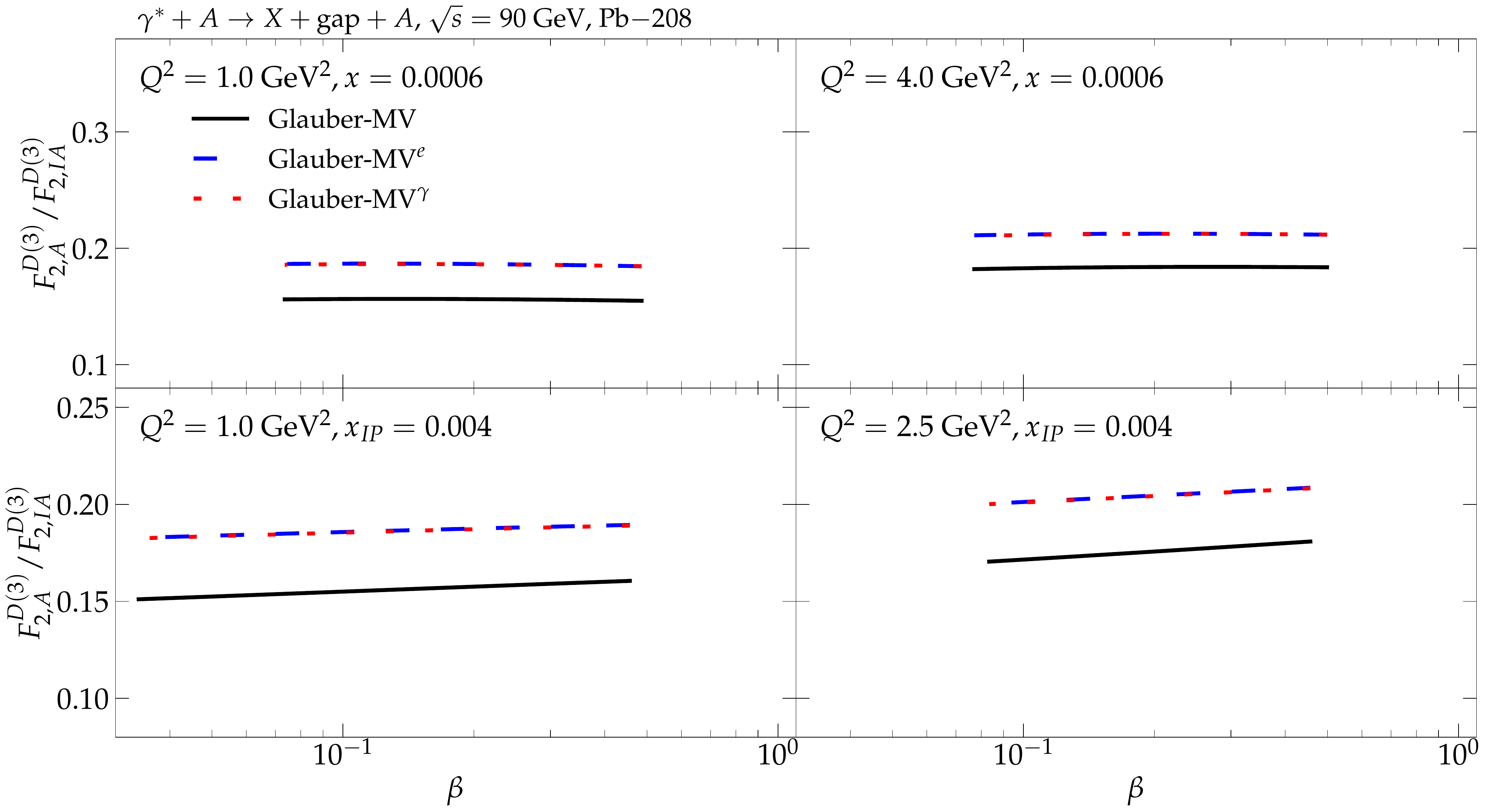}
    \caption{Nuclear modification ratio $F_{2,A}^{D(3)}/F_{2,IA}^{D(3)}$ as a function of $\beta$, when either $\xbj$  (first row) or $\xpom$ (second row) is kept fixed and at different $Q^2$. Only the results with $\beta<0.5$ are shown.}  
    \label{fig:IA_comparison}
\end{figure*}

The nuclear effects can be quantified by comparing the nuclear cross sections to the ones obtained in the impulse approximation.
The impulse approximation corresponds to including the effect of the nuclear geometry (form factor) that controls the $t$ distribution in diffractive scattering, but no other nuclear effects. Thus any deviation from the impulse approximation result in our calculation can, in the dipole picture, be attributed to enhanced saturation effects in nuclei.

In the impulse approximation the diffractive $\gamma$A cross section can be expressed in terms of the diffractive proton cross section at $t=0$ and the nuclear form factor as 
\begin{equation}
    \sigma_{D,IA}^{\gamma^* A} =   \frac{\dd\sigma^{\gamma^*p}_D (|t| = 0)}{\dd{|t|}} \ \Phi_A.
\end{equation}
The nuclear form factor integrated over the squared momentum transfer $-t=\Deltat^2$ reads
\begin{equation} \label{eq:IA_form_factor}
    \begin{aligned}
        \Phi_A &= A^2\int_0^{\infty} \dd{|t|} \left|\int \dd[2]{\bt} e^{-i\bt\cdot {\bf \Delta}}T_A(\bt)\right|^2\\  
        & = 4\pi A^2 \int \dd[2]{\bt} T_A^2(\bt).
    \end{aligned}
\end{equation}
We note that the impulse approximation in practice corresponds to using $b_\mathrm{cut}=0$ in \cref{eq:dipole_nucleus_diffractive_largeb}, i.e. always using a scaled dipole-proton scattering amplitude when calculating diffractive dipole-nucleus interaction. In terms of the diffractive dipole-proton scattering amplitude the diffractive dipole-nucleus cross section in the impulse approximation reads

\begin{equation}
    \label{eq:impulse_approx_qq}
    \sigma_{D,IA}^{q\Bar{q}A} = \frac{\sigma_0^2\mathcal{N}_D(r,Y,Y_0)}{4} A^2\int \dd[2]{\bt} T_A^2(b).
\end{equation}
This can be used in \cref{eq:dip_factorization_diff} to calculate impulse approximation results for the $\gamma^*A$ scattering. Note that the impulse approximation only involves the $t$-differential proton cross section. As a consequenceit can  be written in terms of $\sigma_0$, not involving the proton shape parameter $\omega$.

The diffractive structure function as a function of $\beta$ normalized by the impulse approximation result is shown in \cref{fig:IA_comparison} both at fixed Bjorken-$x$ and at fixed $\xpom$. We will refer to this ratio as the nuclear suppression factor, and with the KL evolution we obtain very strong suppression $\sim 0.15\dots 0.21$ in our chosen kinematics which are accessible at the EIC. The ratios obtained using the MV$^e$ and MV$^\gamma$ parametrizations are in practice identical, and a slightly larger suppression is predicted using the MV fit. This can be compared to predictions for the (much weaker) nuclear suppression  in inclusive hadron production in proton-nucleus collision at the LHC shown in Ref.~\cite{Lappi:2013zma}, where identical  suppression factors are obtained with MV$^e$ and MV$^\gamma$ fits, with slightly weaker suppression obtained with the MV parametrization.

\begin{figure}[tb]
    \centering
    \includegraphics[width=\columnwidth,height=\textheight,keepaspectratio]{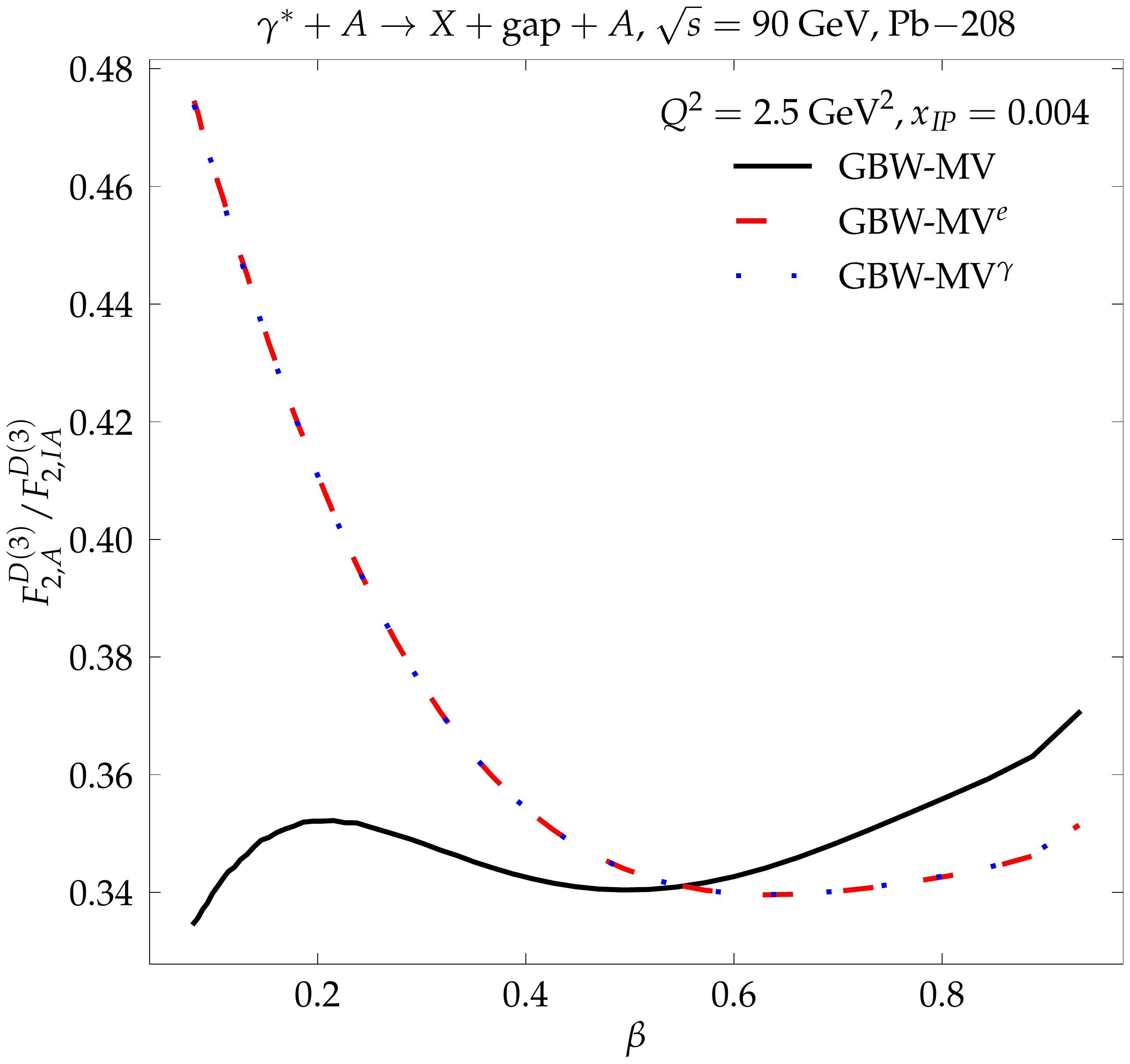}
    \caption{Nuclear modification ratio $F^{D(3)}_{2,A}/F^{D(3)}_{2,IA}$ as a function of $\beta$ using the GBW approach. }
    \label{fig:GBW_IA}
\end{figure}

The  suppression obtained for the diffractive structure functions in the KL approach is much stronger than what is obtained from the GBW setup, which gives $\sim 0.34 \dots 0.48$ at the same kinematics as shown in \cref{fig:GBW_IA}. This strong suppression in the KL approach can again be explained by noticing that the KL evolution modifies the anomalous dimension of the diffractive scattering cross section: the scaling changes as $\mathcal{N}_{D,A}\sim [r^2 Q_{s,A}^2(Y_0,b)]^{2\gamma_c} \to \left[r^2 Q_{s,D(A)}^2(Y,Y_0,b)\right]^{\gamma_c}$. Convoluting with the squared photon wave functions (see the previous section), the nuclear suppression factor at the cross-section level from the KL approach eventually scales as
\begin{equation}
    \label{eq:ratio_estimation_KL}
    \left(\frac{\sigma_{D}^{\gamma^*A}}{  \sigma_{IA,D}^{\gamma^*A} } \right)_{\rm KL} \sim \frac{\int \dd[2]{\bt} \left(\frac{Q_{s,D(A)}^2(b)}{Q^2}\right)^{\gamma_c}}{\sigma_0^2 A^{4/3}\left(\frac{Q_{s,D(p)}^2}{Q^2}\right)^{\gamma_c}} \sim A^{-\frac{1}{3}-\delta(\gamma_c)}\sigma_0^{\gamma_c-2},
\end{equation}
where $\delta(\gamma_c) \approx 0.11$ for $\gamma_c \approx 0.85$, using $Q_{s,A}^2 \sim \sigma_0 A T_A(b)$. A similar evaluation applied for the $q\Bar{q}$ contribution leads to
\begin{equation}
    \label{eq:ratio_estimation_GBW}
    \left(\frac{\sigma_{D}^{\gamma^*A}}{  \sigma_{IA,D}^{\gamma^*A} } \right)_{{\rm GBW}-q\Bar{q}} \sim \frac{\int \dd[2]{\bt} \left(\frac{Q_{s,A}^2(b)}{Q^2}\right)}{\sigma_0^2 A^{4/3}\left(\frac{Q_{s,p}^2}{Q^2}\right)} \sim A^{-\frac{1}{3}}\sigma_0^{-1}.
\end{equation}
We can obviously see that the latter is less suppressed than the former. Furthermore, it is interesting to recall that, while the large dipoles close to the inverse saturation scales dominate the $\br$-integration in the GBW approach, the dominant contribution in the KL approach comes from the smaller dipoles $r\sim 1/Q$. Resummation, which is important at low-$\beta$, leads to a stronger nuclear suppression, while the effect of the non-linear saturation region is diminished!     
As a side note: this effect depends on the fact that we are starting the evolution for both protons and nuclei at the same rapidity where the nuclear saturation scale is larger than the proton one. If one were to start at the same value of $\qs$, i.e. at a higher rapidity for protons than nuclei, the effect would be different.  

The suppression factor calculated from the KL approach is almost independent of $\beta$ at fixed $x$, and decreases very slowly with decreasing $\beta$ at fixed $\xpom$. The weak $\beta$-dependence could be understandable by noticing that, in the KL evolution, both $Q_{s,D(A)}^2$ and $Q_{s,D(p)}^2$ have the same dependence on $Y_0$ and on $Y$, and the former dependence is very mild as mentioned in the previous section. Hence, the nuclear suppression ratio would be almost flat in $\beta$, see~\cref{eq:ratio_estimation_KL}. A weak-$\beta$ variation, particularly when $\xpom$ is kept fixed, is due to the subleading behavior when including also other possible factors in addition to the leading scaling factor $(r^2Q_{s,D}^2)^{\gamma_c}$ in the solutions to the KL equation. When keeping $\xpom$ (or equivalently $\ygap$) fixed, a similar weak $\beta$-dependence should be observed for the $q\Bar{q}$ and $q\Bar{q}g$ components of the GBW result  (\cref{eq:gbw_qq_L,eq:gbw_qq_T,eq:gbw_qqg}) separately.
However, the  sum of the $q\Bar{q}$ and $q\Bar{q}g$ contributions has a stronger $\beta$-dependence, since the nuclear modification of these two components is different, and their relative weight in the cross section has a significant dependence  on $\beta$. 

\begin{figure}[tb]
    \centering
    \includegraphics[width=\columnwidth,height=\textheight,keepaspectratio]{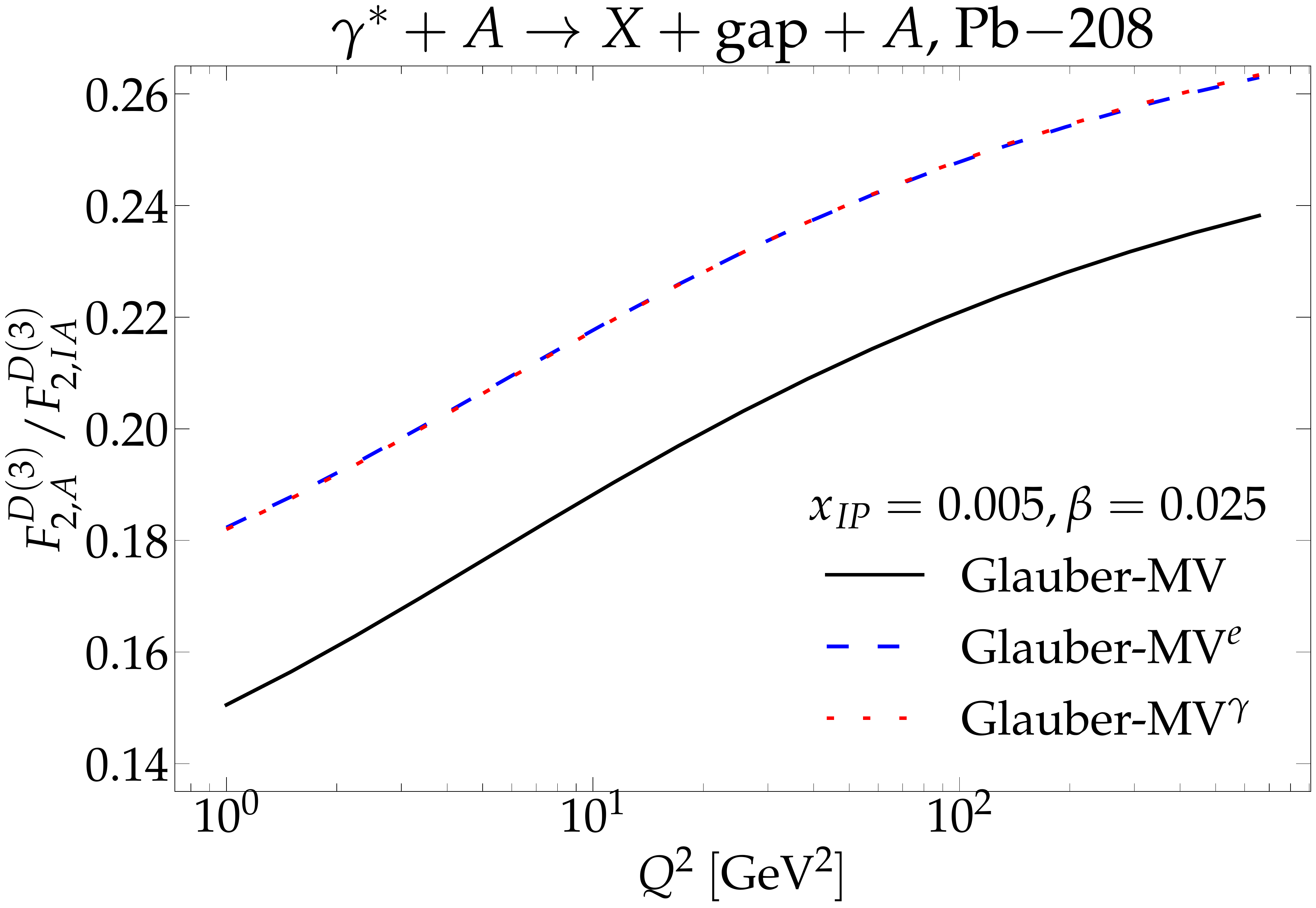}
    \caption{The nuclear modification factor $F_{2,A}^{D(3)}/F_{2,IA}^{D(3)}$ a function of the virtuality $Q^2$. }
    \label{fig:IA_Q2}
\end{figure}
\begin{figure*}[tb]
    \centering
    \includegraphics[width=\textwidth,height=\textheight,keepaspectratio]{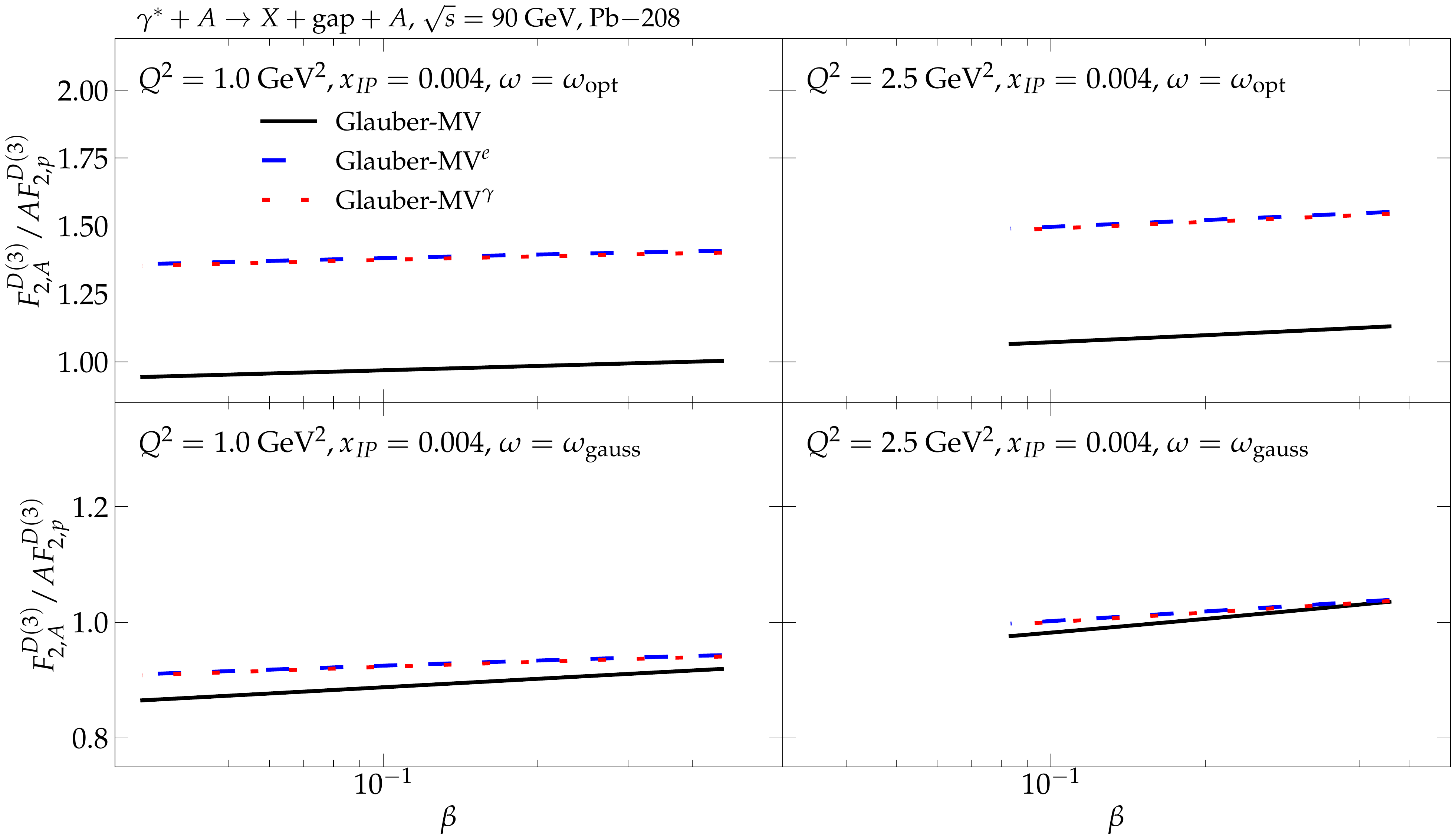}
    \caption{Diffractive structure function ratio $F_{2,A}^{D(3)}/AF_{p,2}^{D(3)}$ as a function of $\beta$ when $\xpom$ is kept fixed. Only the results with $\beta<0.5$ are shown. The results in the first row use the optimal values $\omega_{\rm opt}$ for the steepness of the proton impact parameter profile (see~\cref{sec:hera_comparisons}), while it is the Gaussian value, $\omega_{\rm gauss} = 1$ in the second row.}
    
    \label{fig:nuclear_modification}
\end{figure*}

\begin{figure}[tb]
    \centering
    \includegraphics[width=\columnwidth,height=\textheight,keepaspectratio]{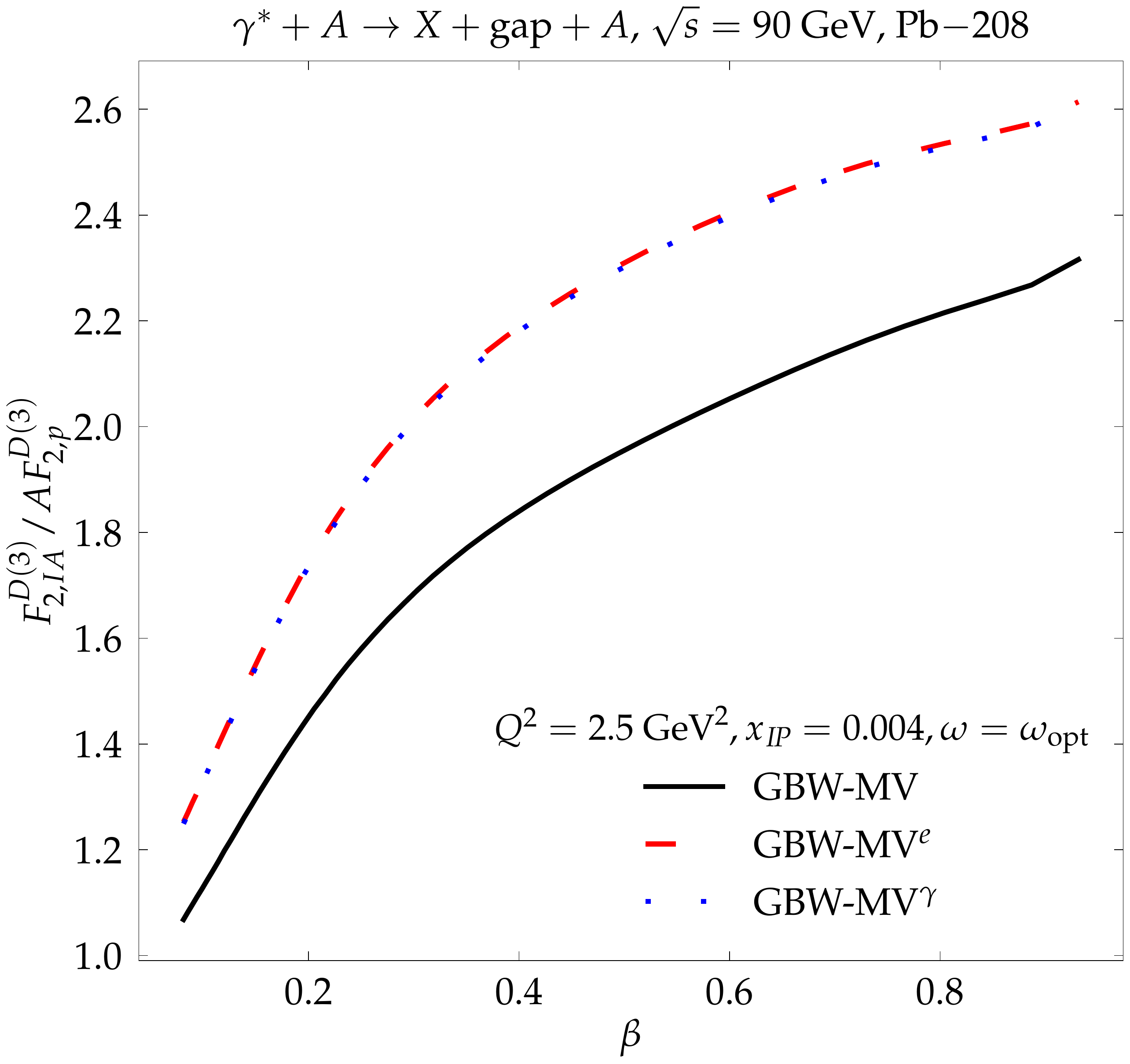}
    \caption{Diffractive structure function ratio $F^{D(3)}_{2,A}/AF^{D(3)}_{2,p}$ as a function of $\beta$ using the GBW approach.}
    \label{fig:GBW_p}
\end{figure}

The virtuality dependence of the 
nuclear suppression factor computed from the KL setup is shown in  \cref{fig:IA_Q2}. As expected a somewhat stronger suppression is obtained towards lower $Q^2$, but even in the large $Q^2\sim 10^3\,\mathrm{GeV}^2$ significant suppression factor $\sim 0.25$ is obtained. This rather weak $Q^2$ -dependence of the suppression can again be understood by considering how the KL evolution changes the anomalous dimension of the diffractive scattering cross section as discussed above.

The nuclear-to-proton diffractive structure function ratio 
$F^{D(3)}_{2,A}/(AF^{D(3)}_{2})$ 
is shown in \cref{fig:nuclear_modification}. 
This ratio again depends weakly on $\beta$, similarly as the case where the impulse approximation is used as a reference.
Note that as the ($t$-integrated) diffractive cross section scales as $\sim A^{4/3}$, this ratio is not normalized such that nuclear effects would vanish in the dilute region. 
The advantage of this structure function ratio is that it depends only on experimentally measurable quantities and there is no need to model the nuclear form factor. It is also directly related to the nuclear modification of the diffractive-to-total cross section ratio, which we will discuss shortly. The normalization factor $A$ (which differs from the parametric $A^{4/3}$ dependence of the nuclear cross section)  allows direct comparisons to earlier works~\cite{Kowalski:2008sa,Aschenauer:2017jsk,AbdulKhalek:2021gbh}.
Unlike the ratio to the impulse approximation, this ratio also depends on the shape of the proton as the normalization of the proton cross section depends  on $\omega$. This dependence on the proton shape is illustrated in \cref{fig:nuclear_modification} by showing the results using both the optimal shapes and the Gaussian shape with $\omega=1$.
The slow increase of this ratio towards larger $\beta$ is qualitatively in agreement with the prediction using the $q\bar{q}g$ component (with or without $q\bar{q}$) presented in Ref.~\cite{Kowalski:2008sa} in the region of $\beta \lesssim 0.1$.

The large $\beta$-region of $\beta>0.1$ has more significant differences between different approaches. In Ref.~\cite{Kowalski:2008sa}, the 
diffractive structure functions were calculated using the GBW formalism. The  IPsat and bCGC models were employed for the $\bt$-dependent proton scattering cross-section, and the nuclear cross-section was obtained directly from the proton case using the Glauber model. For comparison, the result using the GBW approach, but with the BK-evolved dipole amplitudes used in this work, is shown in \cref{fig:GBW_p}. One can see that it produces a rather different prediction from Ref.~\cite{Kowalski:2008sa}. In particular we predict a much larger cross section ratio in the large-$\beta$ region, and additionally in this regime the two calculations have slightly different $\beta$ dependences. These differences can be understandable since the two calculations use different setups for both the scattering off protons and nuclei. Furthermore, the Gaussian profile was used in the cited reference for the proton impact parameter dependence, while in the current calculation, we use the significantly steeper shapes as constrained by HERA data. Note also that our results are closer to prediction using the bCGC set-up than the IPsat one, as the former uses a parameterization for the dipole cross-section based on the solutions to the BK evolution.         

With these dipole amplitude-related differences between results in the GBW formulation in mind, let us then return to the differences between the KL and GBW formalisms.  Comparing the KL result in \cref{fig:nuclear_modification} (the top right panel) to the GBW formalism results in  \cref{fig:GBW_p} and in Ref.~\cite{Kowalski:2008sa}, there is a clear difference in the $\beta$-dependence  in the region of $\beta>0.1$. For the same dipole amplitude (compare the top right panel in \cref{fig:nuclear_modification} to \cref{fig:GBW_p}), the GBW result predicts a larger nuclear enhancement than our present KL approach. Independently of the dipole amplitude, the $\beta$-dependence of the nuclear enhancement is stronger in the GBW approach than in the KL result.
We emphasize again, however, that the KL approach is not fully reliable in the $\beta \gtrsim 0.1$ case. In the  large-$\beta$ regime, the $q\bar{q}$ component dominates, with $F^{D(3)}_2\sim N^2(\xpom)$, and the GBW result treats the kinematics of the small-$M_X$ $q\bar{q}$ state more accurately than the KL approach.

\begin{figure*}[tbp!] 
    \centering
    \includegraphics[width=\textwidth,height=\textheight,keepaspectratio]{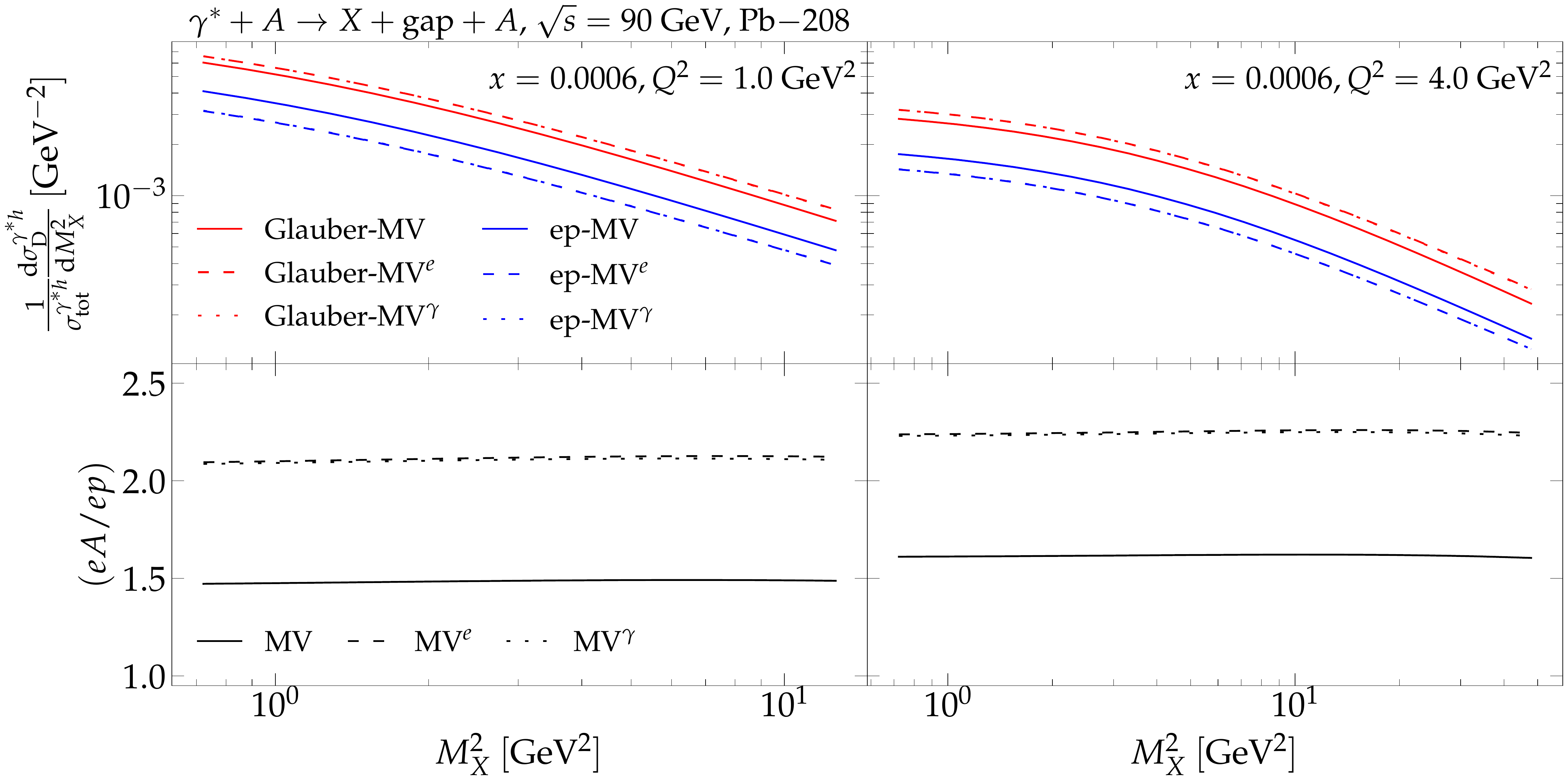}
    \caption{Mass spectra normalized by the total cross sections for $^{208}$Pb and proton (upper row), together with their ratios (lower row) in the corresponding fits, for two values of $Q^2$ at $\xbj=6\times10^{-4}$. The optimal values $\omega_{\rm opt}$ are used for the steepness of the proton impact parameter profile.
    }
    \label{fig:mass_spectra_A_2}
\end{figure*}

Finally we study the diffractive-to-total cross section ratio, as the non-linear nuclear effects are expected to enhance the diffractive cross section relative to the inclusive one~\cite{Kowalski:2007rw}. This ratio as a function of $M_X^2$, and the double ratio
\begin{equation}
    \frac{\rm eA}{\rm ep}  \equiv \left[\frac{1}{\sigma_\mathrm{tot}^{\gamma^*A}}\frac{\dd\sigma_{D}^{\gamma^*A}}{\dd{M_X^2}}\right] {\Big /} \left[\frac{1}{\sigma_\mathrm{tot}^{\gamma^*p}}\frac{\dd\sigma_{D}^{\gamma^*p}}{\dd{M_X^2}}\right]
\end{equation}
are shown in \cref{fig:mass_spectra_A_2}. 
This ratio can also be seen as the nuclear-to-proton diffractive structure function ratio $F^{D(3)}_{2,A}/(AF^{D(3)}_{2})$ divided by the nuclear-to-proton inclusive structure function ratio. A generic feature of gluon saturation is that the fraction of diffractive events in the total cross section should increase when going from protons to nuclei, i.e. the double ratio should be larger than unity. This can be contrasted with the prediction of leading twist shadowing, which would predict a double ratio significantly below one~\cite{Accardi:2012qut}. Thus, this observable is one of the clearest experimental signals for saturation at the EIC.

The result in  \cref{fig:mass_spectra_A_2} confirms that the double ratio is significantly larger than unity.
Again the predictions obtained using the MV$^e$ and MV$^\gamma$ fits are practically identical,  and  a clear nuclear enhancement of $50\%\dots100\%$ is predicted depending on the applied fit. This enhancement is stronger than the GBW prediction shown in Ref.~\cite{Accardi:2012qut}, which can be explained by noting that the double ratio again depends on the proton shape parameter $\omega$, and in this analysis, we indeed have a steeper proton profile rather than the Gaussian shape. The almost-flat behavior of the mass spectrum of the double ratio again resembles the $\beta$ spectrum of the above-mentioned nuclear modification ratios for the diffractive structure function.

\section{Conclusions}
\label{sec:conclusions}

We have presented the first calculation of diffractive cross sections in the HERA kinematics describing the mass dependence by solving the perturbative Kovchegov-Levin (KL) evolution equation\footnote{Note that in Ref.~\cite{Contreras:2018adl}, the authors could  describe rather well the HERA combined data using their analytical solution to the leading-order KL equation in the double-log region.}. Predictions for the future EIC measurements with nuclear targets are also presented. The non-perturbative initial condition for the small-$x$ and high-$\beta$ evolutions is constrained by the HERA structure function data, and the only remaining free parameter describing the shape of the proton (and controlling the overall normalization) is determined from the large-$\beta$ diffractive cross section data. 

Given this input, we find a good description of the precise HERA diffractive structure function and reduced cross section data. The HERA data is found to prefer proton density profiles that are steeper than the commonly-used Gaussian profile. Although in the HERA energy range it is not possible to reach very low $\beta$ (high $M_X^2$) kinematics where the KL evolution dynamics dominates, we find that already  a small amount of KL evolution in the HERA kinematics has a significant effect on the $Q^2$ dependence of the diffractive cross sections. Similarly the KL evolution dynamics results in a very large nuclear suppression for diffractive cross sections in the EIC kinematics in reference to the impulse approximation. The predicted suppression is significantly stronger than what is obtained considering only the fixed photon Fock states $q\bar q$ and $q\bar q g$, i.e. without resumming multiple gluon emissions as is done with the KL evolution. This demonstrates that both the current HERA data and especially the future EIC measurements with nuclei can be used to probe  KL evolution dynamics.

In the future it would be important to more smoothly combine the small-$\beta$ resummation with (LO  and NLO) calculations with more accurate kinematics at high $\beta$. It would also be interesting to simultaneously address the $t$-dependence of exclusive vector meson production and the shape of the proton in inclusive diffraction. This would pave the way towards a more global analysis of inclusive and diffractive deep inelastic scattering cross sections in the dipole picture.

\begin{acknowledgements}
This work was supported by the Academy of Finland, the Centre of Excellence in Quark Matter (project 346324)   and projects 338263 and 346567 (H.M), and  321840 (T.L). This work was also supported under the European Union’s Horizon 2020 research and innovation programme by the European Research Council (ERC, grant agreement No. ERC-2018-ADG-835105 YoctoLHC) and by the STRONG-2020 project (grant agreement No. 824093). The content of this article does not reflect the official opinion of the European Union and responsibility for the information and views expressed therein lies entirely with the authors. 
\end{acknowledgements}

\appendix

\bibliographystyle{JHEP-2modlong.bst}
\bibliography{refs}

\end{document}